\documentclass[longauth,traditabstract]{aa}

\usepackage[T1]{fontenc}

\usepackage{graphicx}
\usepackage{txfonts}
\usepackage{comment}
\usepackage{lscape}
\usepackage{longtable}
\usepackage{amsmath}	
\usepackage{amssymb}	
\usepackage{gensymb}

\usepackage{hyperref}
\hypersetup{breaklinks=true,colorlinks=true,linkcolor=blue,citecolor=blue,urlcolor=blue}

\usepackage{natbib,twoopt}
\bibpunct{(}{)}{;}{a}{}{,}             

\usepackage[usenames]{color}                                   

\begin{document}

\defcitealias{cortijo17}{CF17}
\defcitealias{gonzalezdelgado2015}{GD15}
\defcitealias{lipari2004}{L04}

\title{The spatially resolved stellar population and ionized 
gas properties in the merger LIRG NGC 2623} 
\subtitle{}

\authorrunning{Cortijo-Ferrero et al.}
\titlerunning{Spatially resolved properties of NGC 2623}

\author{
C. Cortijo-Ferrero\inst{1},
R. M. Gonz\'alez Delgado\inst{1},
E. P\'erez\inst{1},
S. F. S\'anchez\inst{2},
R. Cid Fernandes\inst{3},
A. L.\ de Amorim\inst{3},
P. Di Matteo\inst{4},
R. Garc\'{\i}a-Benito\inst{1}, 
E. A. D. Lacerda\inst{3},
R. L\'opez Fern\'andez\inst{1},
C.Tadhunter\inst{5},
M. Villar-Mart\'\i n\inst{6}
and
M. M. Roth\inst{7}
}

\institute{Instituto de Astrof\'isica de Andaluc\'ia (CSIC), PO 
Box 3004 18080 Granada, Spain. (\email{clara@iaa.es})
\and
Instituto de Astronom\'i a,Universidad Nacional Auton\'oma de M\'exico, A.P. 70-264, 04510 M\'exico D.F., Mexico
\and
Departamento de F\'{\i}sica, Universidade Federal de Santa Catarina, P.O. Box 476, 88040-900, Florian\'opolis, SC, Brazil
\and
Observatoire de Paris, GEPI, Observatoire de Paris, PSL Research University, CNRS, Univ Paris Diderot,
Sorbonne Paris Cité, Place Jules Janssen, 92195 Meudon, France
\and
Department of Physics and Astronomy, University of Sheffield, Sheffield S3 7RH, UK
\and
Centro de Astrobiolog\'ia (INTA-CSIC), Carretera de Ajalvir, km 4, E-28850 Torrej\'on de
Ardoz, Madrid, Spain
\and
Leibniz-Institut für Astrophysik Potsdam (AIP), An der Sternwarte 16, 14482 Potsdam, Germany
}

\date{Jun 2017}


\abstract{We report on a detailed study of the stellar populations and ionized gas properties 
in the merger LIRG NGC 2623, 
analysing optical Integral Field Spectroscopy from the CALIFA survey 
and PMAS LArr, multiwavelength HST imaging, and OSIRIS narrow band 
H$\alpha$ and [NII]$\lambda$6584 imaging. The spectra were processed with 
the $\textsc{starlight}$ full spectral fitting code, 
and the results compared with
those for two early-stage merger LIRGs (IC 1623 W and NGC 6090), together 
with CALIFA Sbc/Sc galaxies. We find that 
NGC 2623 went through two periods of increased star formation (SF), 
a first and widespread episode, traced by intermediate-age stellar 
populations ISP (140 Myr--1.4 Gyr), and a second one, traced by young 
stellar populations YSP ($<$140 Myr), which is concentrated in the central
regions ($<$1.4 kpc). Our results are in agreement with 
the epochs of the first peri-center passage ($\sim$200 Myr ago) 
and coalescence ($<$100 Myr ago) predicted by dynamical models, 
and with high resolution merger simulations in the literature, 
consistent with NGC 2623 representing an evolved 
version of the early-stage mergers.
Most ionized gas is concentrated within $<$2.8 kpc, 
where LINER-like ionization and high velocity dispersion 
($\sim$220 km/s) are found, consistent with the previously reported 
outflow. As revealed by the highest resolution OSIRIS and HST data, 
a collection of HII regions is also present in the plane of the 
galaxy, which explains the mixture of ionization mechanisms 
in this system. It is unlikely that the outflow in NGC 2623 will escape 
from the galaxy, 
given the low SFR intensity ($\sim$0.5 M$_{\odot}$yr$^{-1}$kpc$^{-2}$), 
the fact that the outflow rate is 3 times lower than the current SFR, and the 
escape velocity in the central areas higher than the outflow velocity.}

\maketitle

\section{Introduction}
\label{1}

NGC 2623 (UGC 04509 = Arp 243; 
L$_{IR}$ = 3.56 $\times$ 10$^{11}$ L$_{\odot}$; e.g., \citealt{sanders2003}) 
is an infrared luminous merger of galaxies 
located at a distance of 80 Mpc. 
It is an advanced merger, in stage IV \citep{veilleux2002}, 
where the progenitor's nuclei have already coalesced. 
The nucleus is extended and quite obscured in the ultraviolet (UV) 
and optical, and becomes very bright and point-like in 
infrared images. Moreover, this galaxy has two 
long and fairly symmetrical
tidal tails, extending up to 20 kpc from the nucleus. 
The galaxy also shows a blue star forming region south of the main body \citep{evans2008}. 
The near-infrared (NIR) surface brightness profile in the nuclear regions follows a 
r$^{1/4}$ law, consistent with a system evolving into an elliptical galaxy 
\citep{wright1990,rothberg&joseph2004}.
A substantial post-starburst population dominates in the 
circum-nuclear region of NGC 2623 
\citep{joy&harvey1987,armus1989,liu&kennicutt1995,shier1996}, while a young 
nuclear starburst is also present \citep{bernloehr1993,shier1996}.
\cite{evans2008} reported that the blue, UV bright star
forming region located 5 kpc south of the nucleus comprises 
$\sim$ 100 star clusters with ages between 1 and 100 Myr. 
This region could be either part of a loop of material associated 
with the northern tidal tail, or nuclear tidal 
debris created during one of the final passages of the progenitor nuclei 
prior to their coalescence.
The bulk of infrared luminosity comes from the obscured 
nuclear regions, and the off-nuclear starburst represents only 1$\%$ 
of the total star formation in NGC 2623.
The ionization process in the nucleus of NGC 2623 is a combination 
of a dusty starburst and shock heating due to an starburst related 
outflow \citep{lipari2004}. 
The outflow has an opening angle 
$\theta$ = 100$\degree$ $\pm$ 5$\degree$ and reaches 
a distance of 3.2 kpc from the nucleus and a velocity of 
V$_{OF}$=(-405 $\pm$ 35) km s$^{-1}$. 
The nucleus hosts an obscured AGN, as determined from the 
hard X-ray spectrum \citep{maiolino2003}
and the presence of the [NeV] 14.3 $\mu$m line 
\citep{evans2008,petric2011}, which contributes to 
9$\%$ of the total nuclear radio emission, being energetically weak 
relative to the starburst population \citep{lonsdale1993}.
\begin{figure*}  
\begin{center}
\includegraphics[width=0.8\textwidth]{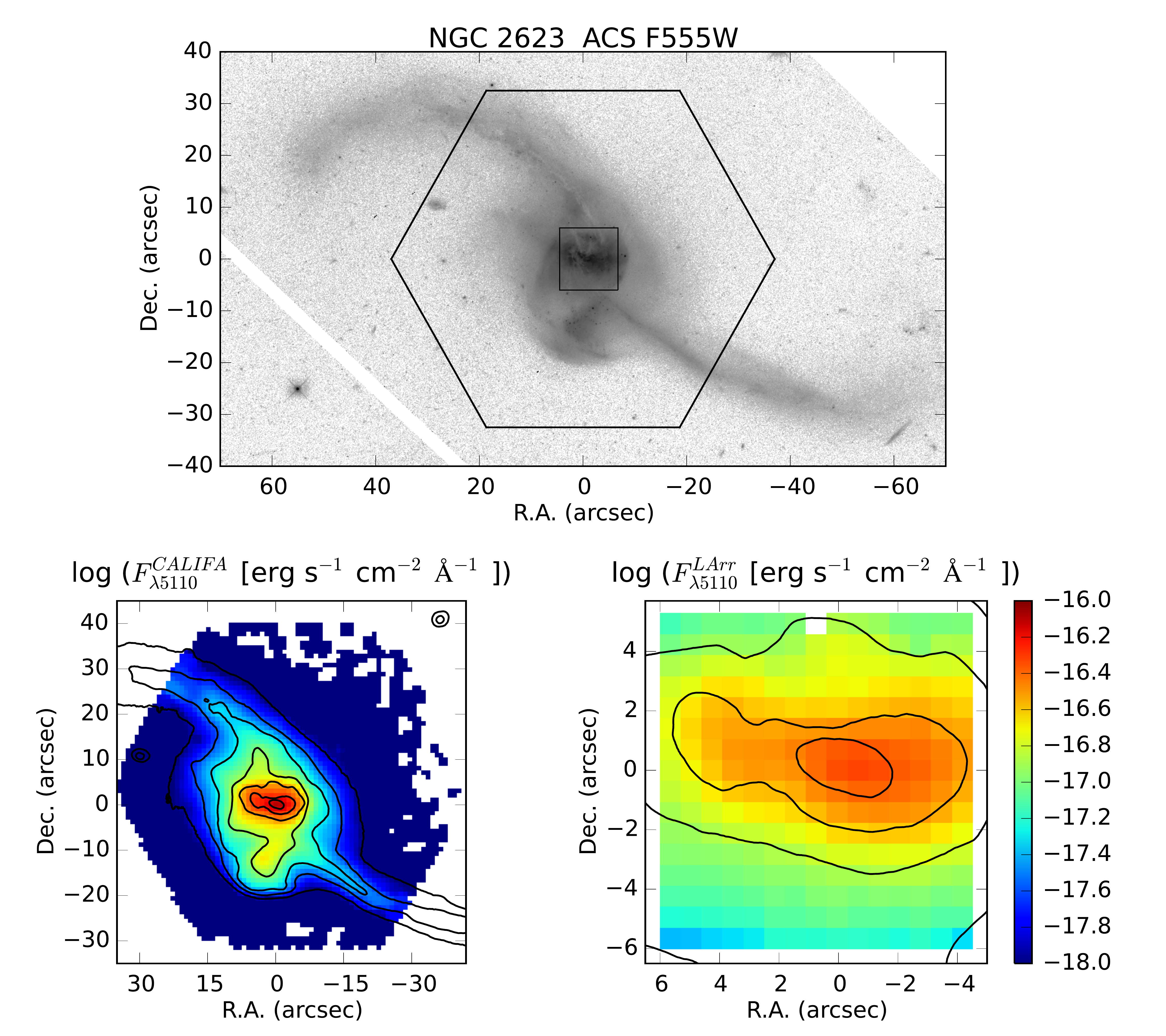} 
\caption{Top panel: HST ACS F555W image of NGC 2623. 
The black rectangle indicates the FoV covered by the LArr IFS, 
while the black hexagon indicates PPaK FoV. 
Lower left panel: continuum flux at 5110 $\rm \AA$ rest-frame (obtained by 
averaging the spectra in the range 5050--5170 $\rm \AA$), F$_{\lambda 5110}$, for the 
CALIFA data. Lower right panel: F$_{\lambda 5110}$ for LArr data.}
\label{Fig_1b}  
\end{center}
\end{figure*}

In this paper, we aim to derive both the stellar 
population properties, and the ionized gas properties of NGC 2623, 
in order to characterize in detail the merger-induced star formation and 
the ionization structure in the final stages of a major merger.
Previous longslit studies in other merging (ultra)luminous infrared 
galaxies (U/LIRGs) have revealed that stellar populations
of $<$ 2 Gyr are present widespread in these systems, while SP $<$ 100 Myr are 
more significant in the nuclear regions of the galaxies \citep{rodzau2009}. 
Although low resolution merger simulations initially 
predicted that most star formation occured in the central 
regions \citep{mihos&hernquist1996,moreno2015}, 
the new high resolution models show that extended star formation 
is also present and important in the early phases of 
mergers \citep{teyssier2010,hopkins2013,renaud2015}. 
In this sense Integral Field Spectroscopy (IFS) is a very promising 
technique, because it can provide relevant information to characterize 
the extent of star formation, and how/when it is produced.
Extended star formation has also been observationally reported 
in many early-stage mergers, mainly in the form of widespread 
star clusters, most of which are located at the intersections between 
progenitors and/or tidal structures \citep{wang2004,elmegreen2006,smith2016}, 
and through stellar population analysis 
relying on IFS (Cortijo-Ferrero et al. 2017, \citetalias{cortijo17} hereafter). 
In fact, the results of the two early-stage merger LIRGs 
reported in \citetalias{cortijo17}, IC 1623 W 
and NGC 6090, will be compared through the paper 
with the merger LIRG NGC 2623.
The advantage of NGC 2623 is that it is a more advanced system, 
in the merger stage, where a triggering of the star formation is 
expected to occur, but at the same time, it also 
keeps a fossil record in the stellar populations of previous 
star formation bursts (i.e. when it was at the early-stage 
merger phase). 
Therefore, NGC 2623 represents an interesting nearby LIRG 
where to study the role that major mergers play in galaxy 
evolution using spatially resolved spectroscopy.

The layout of the paper is: Sect. 2 describes the observations and 
data reduction process. In Sect. 3 we apply the fossil record method to 
analyze the stellar continuum and 
derive the spatially resolved stellar population properties: stellar mass and 
stellar mass surface density, $\mu_{\star}$; stellar extinction, A$_{V}$; 
luminosity weighted mean age, $\langle \log age \rangle_{L}$; and the 
contributions to luminosity and mass of young, intermediate and 
old stellar populations. Section 4 presents results on the ionized gas 
emission, focussing on the morphology, nebular A$_{V}$, 
the ionization conditions and velocity dispersion. 
We discuss the results in 
Sect. 5; and Sect. 6 presents the conclusions. 

Throughout the paper we assume a flat cosmology with
$\Omega_{M}$= 0.272, $\Omega_{\Lambda}$ = 0.728, and 
H$_{0}$ = 70.4 km s $^{-1}$ Mpc $^{-1}$ 
(WMAP, seven year results). 
For the NGC 2623 redshift (z=0.018509), this results in a 
distance of 80.0 Mpc. 
At this distance 1$\tt{''}$ corresponds to 0.390 kpc.
\begin{figure*}  
\begin{center}
\includegraphics[width=0.38\textwidth]{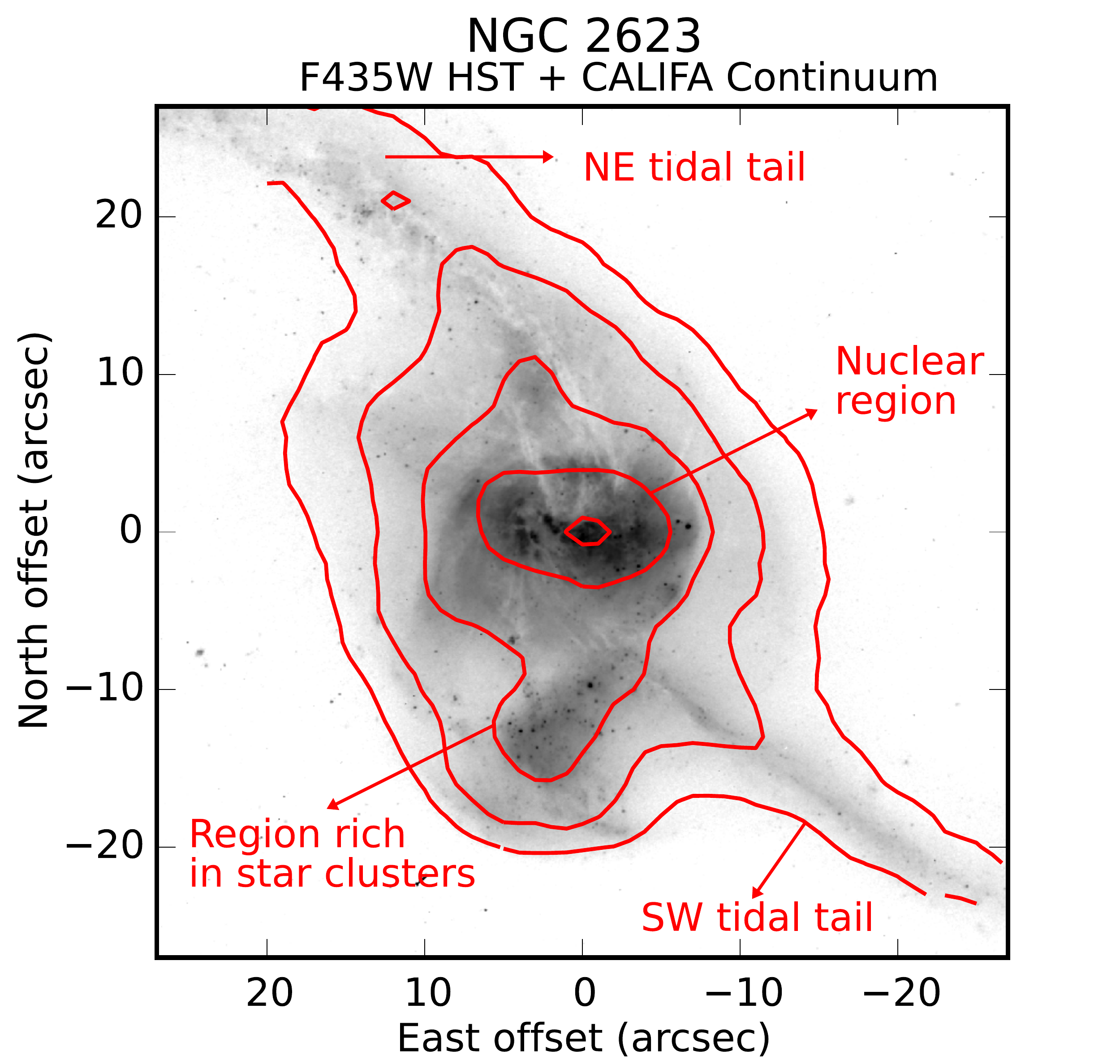} 
\includegraphics[width=0.6\textwidth]{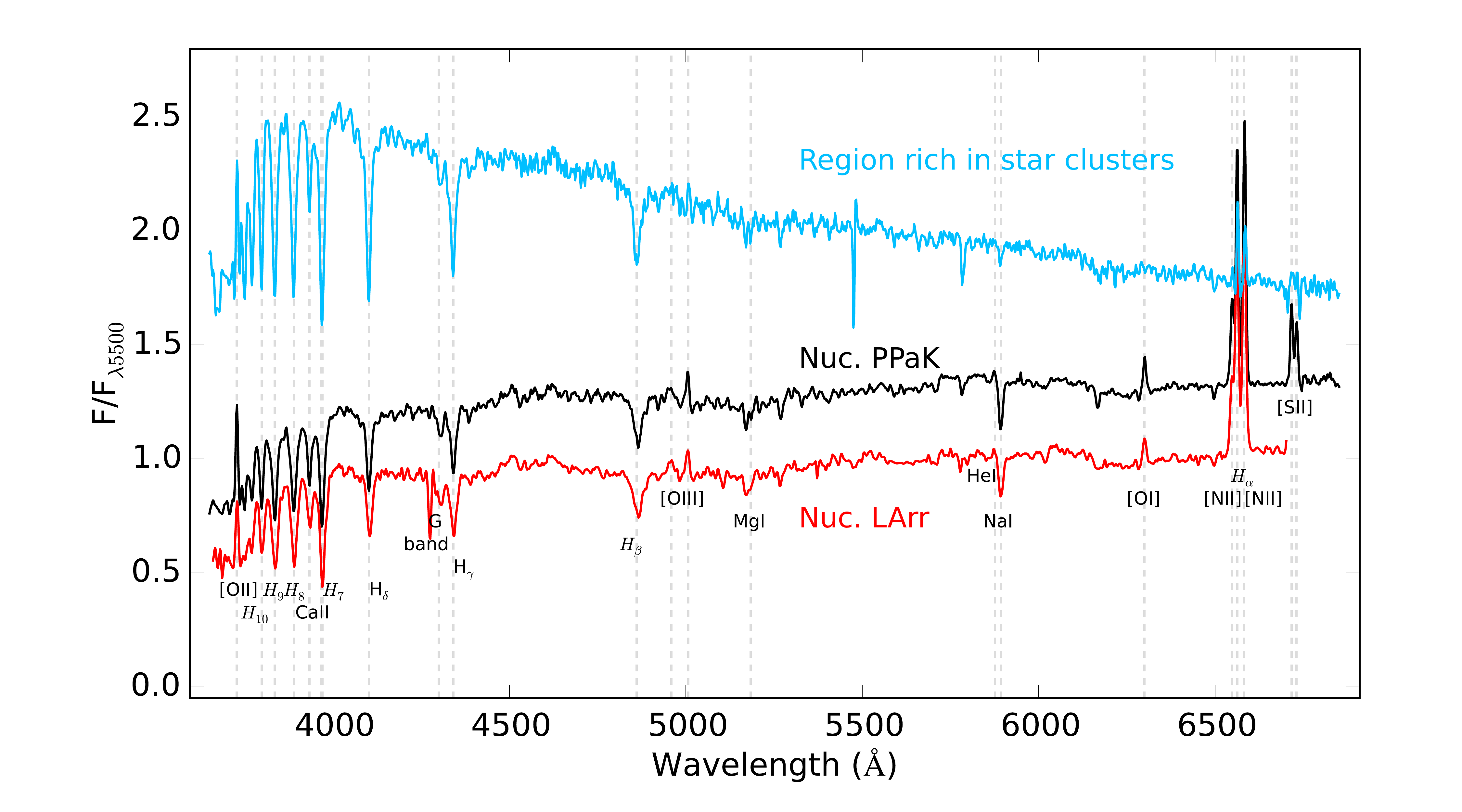} \\
\caption{Left panel: HST F435W continuum image of 
NGC 2623, with the average continuum from CALIFA IFS 
superimposed using red contours. Several regions of the galaxy are labelled.
Right panel: spectra from different regions extracted in circular 
apertures of 2$\tt{''}$ radius, in black the PPaK nuclear spectrum, 
in red the LArr nuclear spectrum, 
and in blue the spectrum from the cluster rich region to the south of the nucleus. 
The spectra are normalized by the flux 
at 5500 $\AA$, and an offset applied to better visualize the differences.}
\label{Fig_1b}  
\end{center}
\end{figure*}


\section{Observations and data reduction}\label{2}
\subsection{Integral Field Spectroscopy}\label{2.1}
The IFS data were taken with the 
Potsdam Multi-Aperture Spectrophotometer 
(PMAS) spectrograph \citep{roth2005} at the 3.5m telescope 
of Calar Alto Observatory (CAHA). The nuclear region 
was observed in the Lens Array (LArr) mode, 
using a spatial magnification of 0.75$\tt{''}$/lens, 
covering a 12$\tt{''}$x 12$\tt{''}$ field of view (FoV).  
The V300 grating was used, providing a 3.2 $\AA$/pixel dispersion and 
covering a wavelength range  3700--7100 $\rm \AA$ with a resolution 
of 7.1 $\rm \AA$ FWHM (full width at half maximum), for a total exposure time of 2.5h.

We also have data from the Calar Alto Integral Field Area (CALIFA) survey 
project \citep{sanchez2012}, covering the main body of NGC 2623, 
and including virtually the entire extent of the system, 
except the outermost parts of the tidal tails.  
The CALIFA observations were conducted using PMAS in its 
PPAK (PMAS fibre PAcK) mode. Fibers in the PPAK bundle have a projected 
diameter on the sky of 2.7$\tt{''}$, covering 
64$\tt{''}$ $\times$ 74$\tt{''}$ FoV. 
The spatial resolution of CALIFA data is 
FWHM$_{PSF}$ = 2.39 $\pm$ 0.26 arcsec \citep{garcia-benito2015}.
Observations were performed using the gratings V500 and V1200, 
with resolutions of 6.3 $\rm \AA$ and 2.3 $\rm \AA$ (FWHM), 
wavelength ranges of 3650--7500 $\rm \AA$ and 3650--4840 $\rm \AA$, 
and total exposure times in each of 0.75h and 1.5h, respectively.
Table 1 summarizes the main characteristics of the IFS data.

\subsubsection{LArr data reduction}\label{2.1.2}
The PMAS LArr data reduction process is 
described in detail in \citetalias{cortijo17}, which should be 
consulted for all the reduction details.
Relative flux calibration was performed using the 
spectrophotometric standard star Feige 34. 
We have flux-recalibrated 
LArr data using HST (WFC435W, WFC555W) and SDSS (g, r-band) photometry,
and have found an average accuracy of the spectrophotometric calibration of 
13 $\%$ $\pm$ 2 $\%$ across the wavelength range covered by our data set.

\subsubsection{CALIFA data reduction}\label{2.1.3}
The raw data were processed through an automatic pipeline, 
and the reduced data cubes made publicly available for the 
community \citep{husemann2013}. 
In our case, we have analysed the V500 combined with the V1200 data, to 
improve the signal in the blue side, and to reduce the effects of  
vignetting. 
The data analyzed in this paper were reduced using the 
CALIFA Pipeline version 1.5. The main reduction steps and properties 
of the reduced data are similar to the LArr ones, and are described in detail 
in \cite{sanchez2012}, based on version 1.2. 
A list of the differences and improvements with respect 
to this earlier version are presented in \cite{husemann2013}.
The combined V1200 + V500 datacubes were processed as described in 
\cite{cidfernandes2013}, and detailed in Section \ref{3.1}. 
\begin{figure}  
\begin{center}
\includegraphics[width=0.5\textwidth]{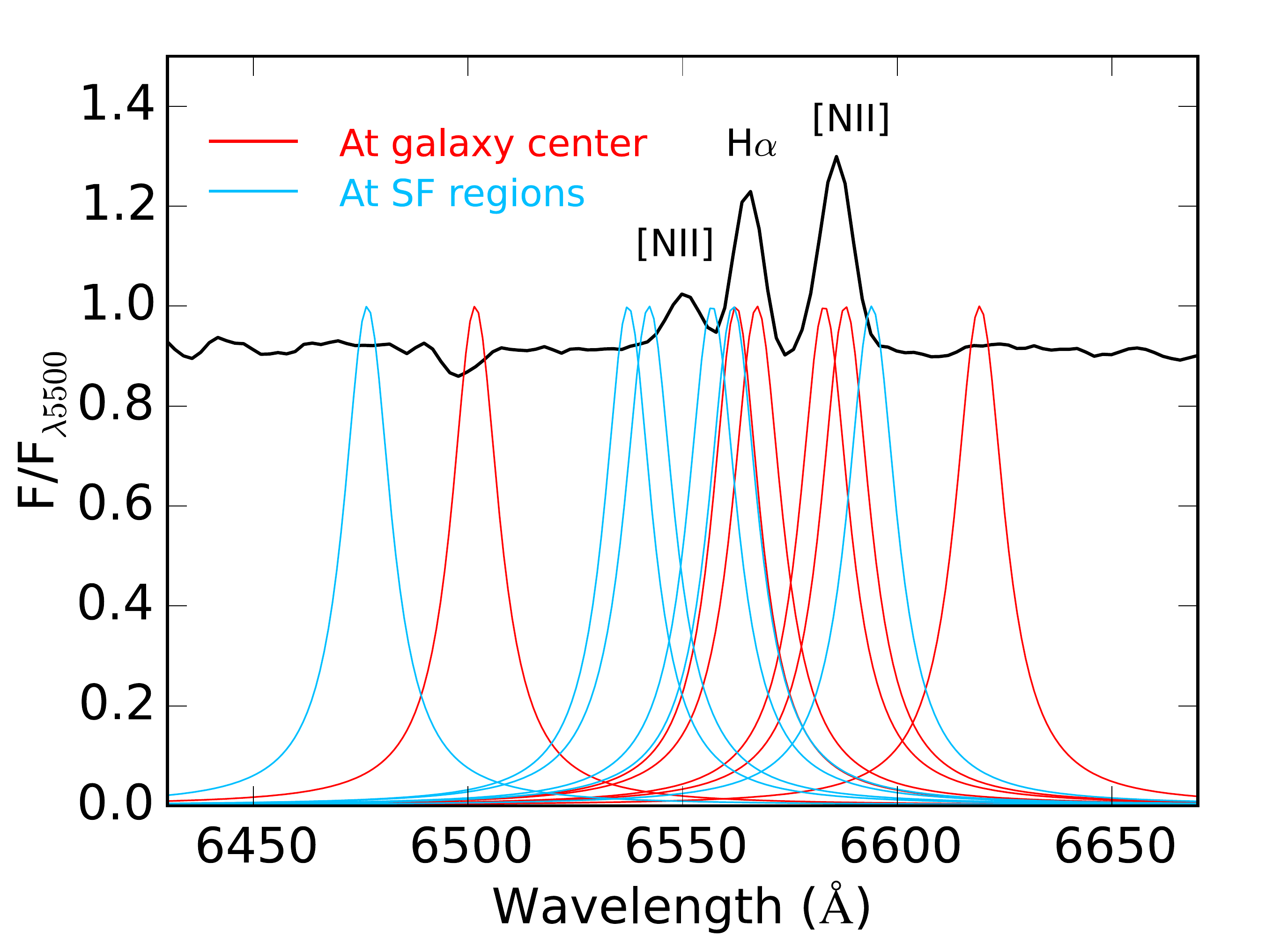} 
\caption{Integrated CALIFA spectrum for NGC 2623 normalized at 5500 $\AA$, 
in the H$\alpha$ + [NII] region. The OSIRIS TF narrow-band filter 
bandpasses at the 
galaxy centre are shown in red, while the positions 
of the filters at the location of the star forming regions (R1 to R4, see
Section \ref{4.7}) are shown in blue.}
\label{Fig_1b}  
\end{center}
\end{figure}

\begin{table*}
\centering
 \caption{Summary of NGC 2623 IFS and TF narrow-band data}
 \label{tab:natbib}
 \begin{tabular}{cccc}
\hline
Data/Instrument         &   LArr-PMAS	                      &    PPaK-PMAS (CALIFA)            &   TF-OSIRIS    \\
\hline
Observation date        &   31/03/2008                        &    20/01/2012	                 &   12/01/2010    \\
Total exposure time (s) & 2.5h (3 x 1200s + 3 x 1800s)        &  0.75h (3 x 900s)                &  1h (6 x 600s)  \\
Field of View           & 12$\tt{''}$ $\times$ 12$\tt{''}$    & 64$\tt{''}$ $\times$ 74$\tt{''}$ & 260$\tt{''}$ $\times$ 520$\tt{''}$ \\
Spatial scale           & 0.75$\tt{''}$/spaxel                & 1$\tt{''}$/spaxel                & 0.25$\tt{''}$/pixel \\
Spatial resolution      & $\sim$ 1.7$\tt{''}$                 & $\sim$ 2.4$\tt{''}$              & $\sim$ 0.9$\tt{''}$  \\
Grating - Spectral resolution & V300 -  7.1$\AA$              & V500 -  6.3$\AA$                 & H$\alpha$+[NII] FWHM$_{TF}$=14$\AA$ \\
\hline  
\hline
 \end{tabular}
\end{table*}

\subsection{OSIRIS Tunable Filters data and reduction}\label{2.2} 
We have taken Tunable Filter (TF) observations with OSIRIS$@$Gran 
Telescopio de Canarias (GTC) at the Roque de los Muchachos Observatory 
in La Palma.
OSIRIS is an imager and spectrograph for the optical wavelength range, located in 
the Nasmyth-B focus of GTC. 
The OSIRIS observations were performed 
using TF imaging mode with a seeing $\sim$ 0.9$\tt{''}$. 
A FoV of 260 $\times$ 520 arcsec square was covered with a spatial scale 
of 0.254 $\tt{''}$/pixel.
The red TF in scanning mode was used, covering H$\alpha$ and [NII]$\lambda$6584 lines 
plus the adjacent continua.
Filters were centreed at $\lambda_{0}$ = 6640, 6700, 6705, 6720, 6725, 
and 6760 $\rm \AA$, with 
a FWHM = 14 $\rm \AA$. These are the wavelength values at the optical centre. 
The wavelength tuning is not uniform over the full FoV of OSIRIS. 
There is a progressive increasing shift to the blue of the central wavelength 
($\lambda_{0}$) as the distance from the optical centre (r) increases. 
The wavelength (in $\rm \AA$) observed with the red TF relative to the optical 
centre changes following the law reported by \cite{gonzalez2014}:
\begin{equation}
\lambda = \lambda_{0} - 5.04 \times r^{2}
\end{equation}
where r is the distance in arcminutes between the
optical centre and the CCD position at which we want to calculate the wavelength.
In Figure 3 we show the integrated CALIFA spectrum for NGC 2623, 
normalized at 5500 $\AA$, 
in the H$\alpha$ + [NII] region. In red we show the OSIRIS TF 
narrow-band filter bandpasses at the galaxy centre, and in blue the position 
of the filters at the location of the star forming regions. 
Note how, due to the greater distance of the star forming 
regions from the optical centre, the [NII] filter is displaced 
into the H$\alpha$ region.
Distances to the optical centre above 2.2$\tt{'}$ and 
below 1.3$\tt{'}$ start to be affected by the shift in wavelength. 
In our case, just the edges of the tidal tails of NGC 2623 are affected, 
while the main body and the clumps in the north tidal 
tail (at 1.65$\tt{'}$) are not.
The object and the calibration star (Feige 34) were
positioned at the same distance from the optical centre and on 
the same CCD chip. 
We obtained three science frames per filter of 300 s exposure time each. 
The main information about TF data is summarized in the right column 
of Table 1.

The OSIRIS data were reduced using standard procedures for CCD imaging 
within the IRAF package. Due to 
the high dark level in the OSIRIS CCDs, 
a series of dark images with the same exposure time 
as the science data were taken. 
To remove the presence of several sky emission rings at the left edges 
of the image we developed a simple Python script that generated 
an artificial background map from the original images.
Finally, a relative flux calibration was performed 
for each filter by comparing the aperture
photometry of Feige 34 measured in OSIRIS data, with
the theoretical value obtained from \cite{oke1990} calibrated
spectra. 

Once we had reduced all the science images, a mean continuum map was 
obtained by averaging the blue (6640 $\rm \AA$) and 
red (6760 $\rm \AA$) continua.
Analogously, we obtained mean H$\alpha$+continuum and 
[NII]+continuum maps by averaging the 6700 $\rm \AA$ and 
6705 $\rm \AA$, and 
6720 $\rm \AA$ and 6725 $\rm \AA$ filter data, respectively. 
By subtracting the mean continuum map from the latter images, 
we obtained the pure H$\alpha$ and [NII] 
emission line maps shown in Section \ref{4}.

\subsection{HST imaging}\label{2.3} 
We downloaded from the Hubble 
Legacy Archive (HLA)\footnote {\url{http://hla.stsci.edu/}} 
multi-wavelength high resolution images taken by the Hubble Space 
Telescope (HST) in several broad-band 
filters from far ultraviolet (FUV) to NIR: 
ACS F140LP, ACS F330W, ACS F435W, ACS F555W, ACS F814W, NICMOS F110W,  
and NICMOS F160W,
covering the main body of NGC 2623. From these we derived 
the star clusters photometry. 
All these images were retrieved in their pipeline reduced form, 
astrometrically corrected and aligned with North up, East left.
Their characteristics are summarized in Table A.1. 

\subsection{An overview of the data}\label{2.4} 
Figure 1 presents an overview of the data, indicating the 
FoV covered by LArr/PPaK over the HST/ACS F555W image, and the IFS 
continuum flux at 5110$\AA$.
In the left panel of Figure 2 we have highlighted, 
over the HST/ACS F435W image and PPaK continuum contours, 
some of the main regions of the galaxy. 
Their spectra are shown in the right panel. 
The nuclear spectra from LArr (black) and from CALIFA data (red) show good
agreement in terms of the spectral continuum shape. They show prominent 
emission lines, as well as the high order Balmer lines in absorption, indicating 
the coexistence of young and intermediate age stellar populations. 
Dust extinction is important in the nucleus, as revealed by the continuum 
rise beyond 6000 $\AA$ \citep{sanchez-almeida2012}. The spectrum of the 
star cluster region to the south is shown in blue, and is consistent 
with a stellar population with an age of a few Myr. Weak H$\alpha$ and [NII] is also
present.

\begin{figure*} 
\begin{center}
\includegraphics[width=0.7\textwidth]{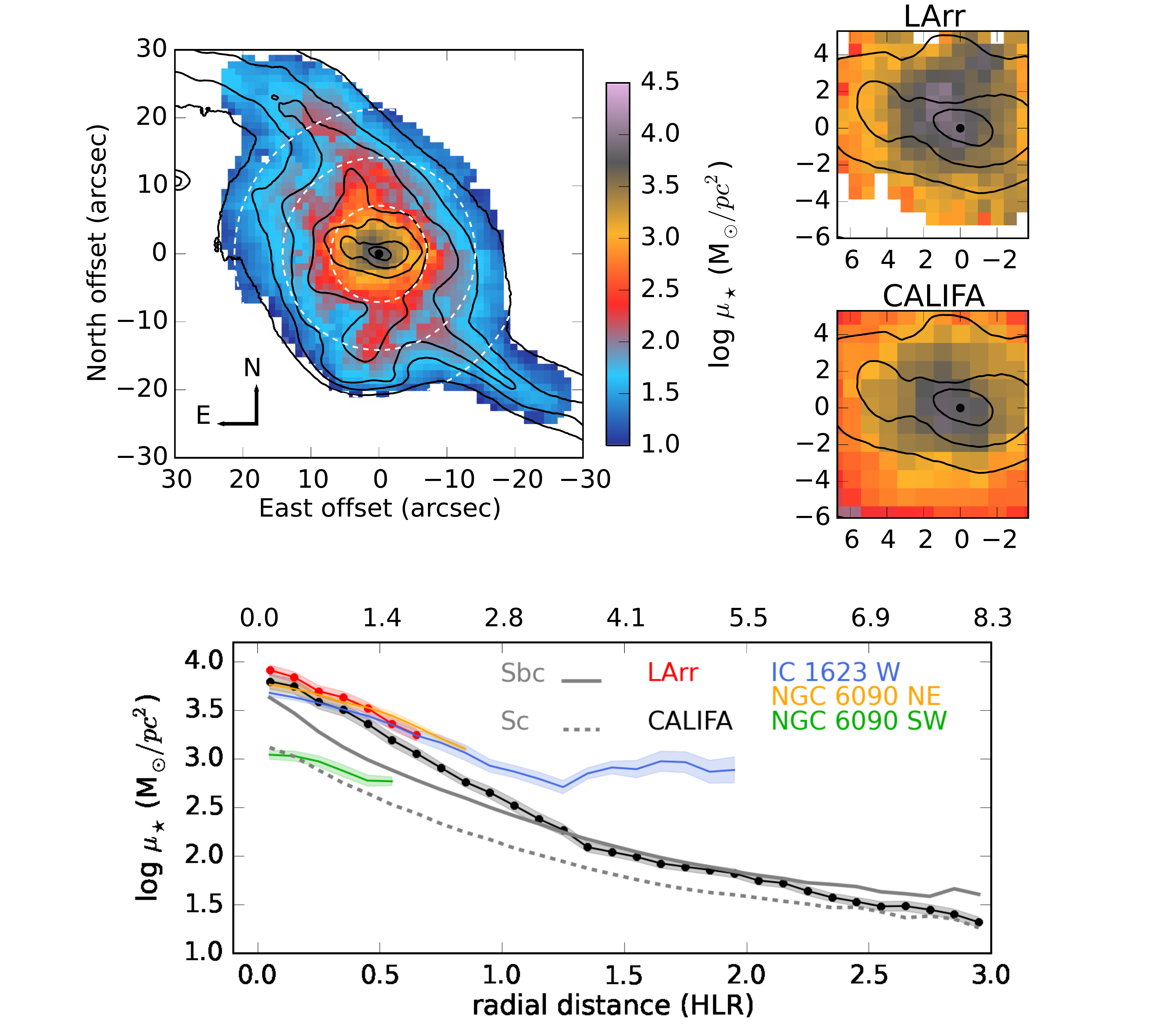}
\caption{Upper left: CALIFA map of the stellar mass surface density ($\mu_{\star}$). 
The white dashed lines indicate the position of 1, 2, and 
3 half-light radii (HLR); 1 HLR is equivalent to 2.8 kpc in 
physical distance. The contours correspond to the smoothed HST F555W image. 
Upper right: LArr map of stellar mass surface density. 
The colour scale is the same as for the CALIFA map. Below is presented a zoom of the CALIFA map in 
the nuclear region, to match the region covered by LArr map. Lower panel: radial
profile of log $\mu_{\star}$ as a function of the radial 
distance in HLR (lower horizontal axis) and kpc (upper horizontal axis), 
in red for LArr and black for CALIFA. 
The uncertainties, calculated 
as the standard error of the mean, are shaded in light red and grey, 
respectively.
For comparison, the grey lines are the profiles from Sbc (solid) and Sc (dashed) 
spiral galaxies from CALIFA \citep{gonzalezdelgado2015}, 
with stellar masses similar to that of NGC 2623. 
We also include the profiles of the early-stage 
mergers IC 1623 W (blue), NGC 6090 NE (orange), 
and NGC 6090 SW (green).}
\label{Fig_4p10}  
\end{center}
\end{figure*}
\begin{figure*}
\begin{center}
\includegraphics[width=0.7\textwidth]{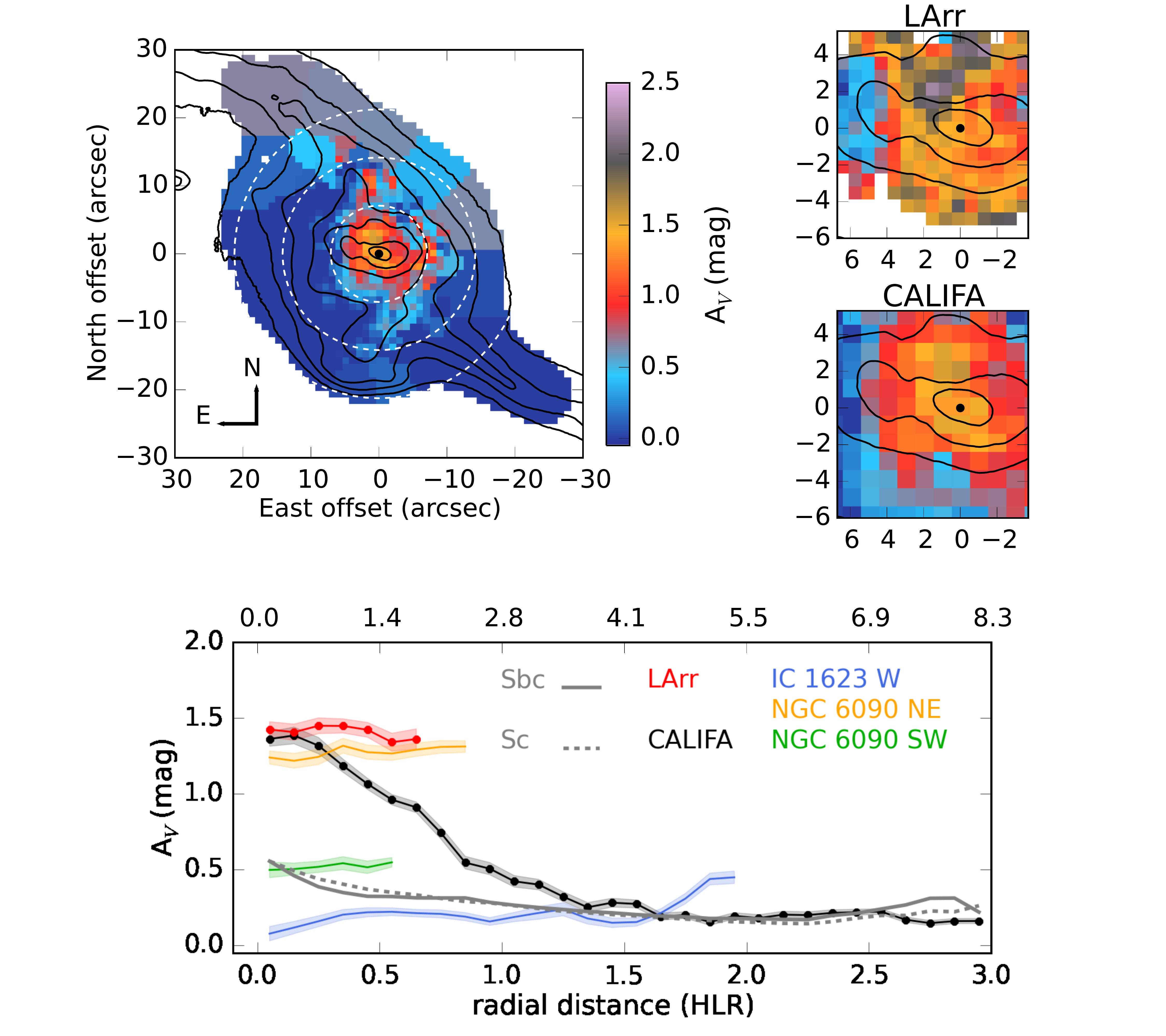}
\caption{As Figure 4 but for stellar dust extinction.}
\label{Fig_4p11}  
\end{center}
\end{figure*}
\begin{figure*} 
\begin{center}
\includegraphics[width=0.7\textwidth]{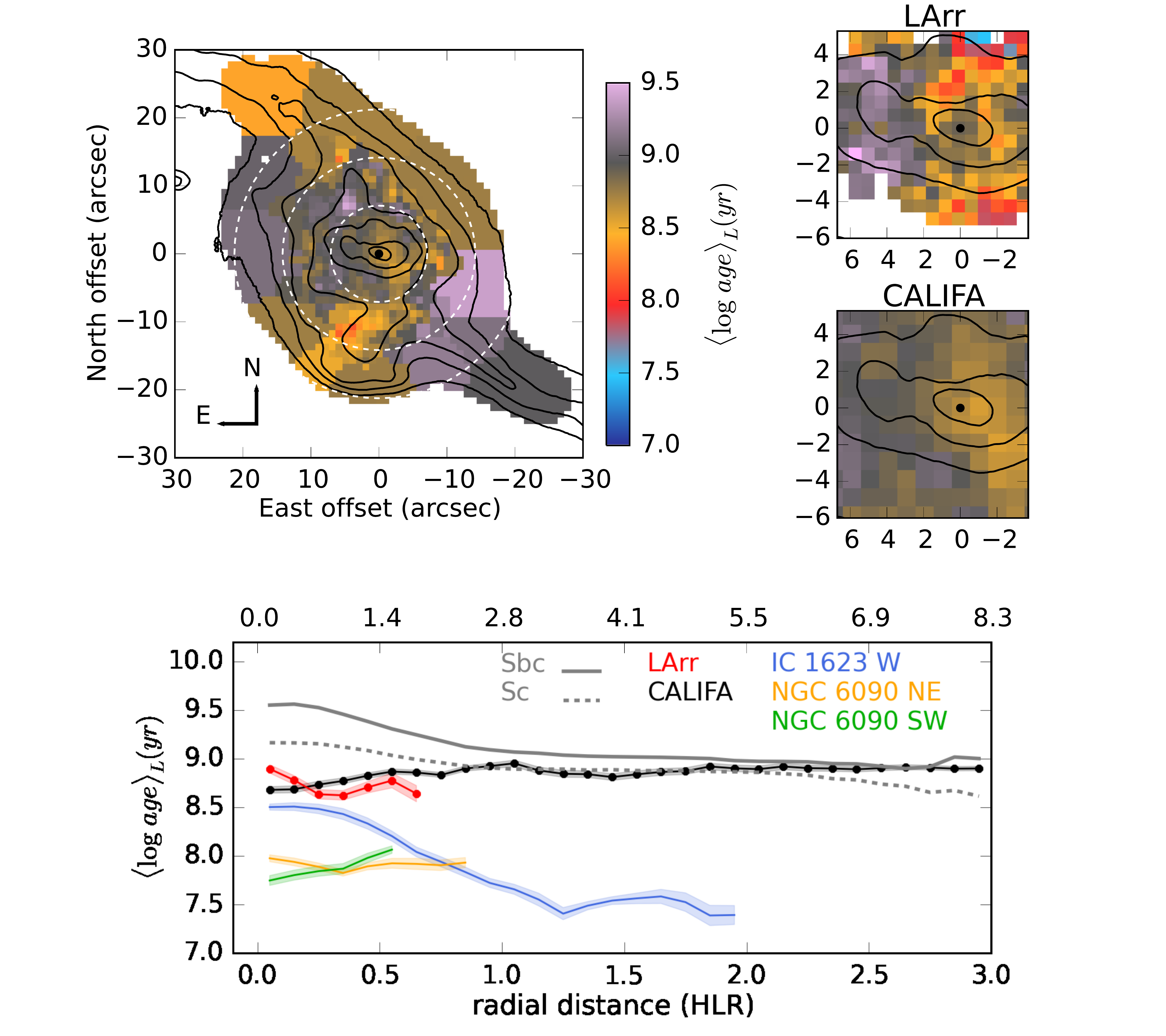}
\caption{As Figure 4 but for mean light weighted stellar ages.}
\label{Fig_4p12}  
\end{center}
\end{figure*}


\section{Stellar populations}\label{3}
In this section we use the stellar continuum shape from the IFS 
to characterize the spatially resolved stellar population 
properties and constrain the star-formation history (SFH) of NGC 2623.
We have also performed an analysis based on the star clusters 
detected in HST images that is presented in Appendix A.

Moreover, in this Section we  compare 
with the radial profiles 
of two control samples of 
70 Sbc and 14 Sc CALIFA spirals \citep{gonzalezdelgado2015}, 
in the same mass range as NGC 2623.

\subsection{Methodology}\label{3.1}
Our method of extracting stellar population information 
from the IFS data cubes is based on the full spectral 
synthesis approach.

For the CALIFA data cube, spectra suitable for a full spectral synthesis 
analysis of the stellar population content were extracted in a way that ensures 
a signal-to-noise ratio $S/N \ge 20$ in a 90 $\AA$ wide region centred at 
5635 $\AA$ (in the rest-frame). When individual spaxels did not meet this 
$S/N$ threshold they were coadded into Voronoi zones \citep{cappellari&copin2003}. 
Pre-processing steps also included spatial masking of foreground/background sources 
and very low $S/N$ spaxels, rest-framing, and spectral resampling. 
The whole process took 
spectrophotometric errors ($e_{\lambda}$) and bad pixel flags ($b_{\lambda}$) 
into account. The spectra were then processed through PyCASSO 
(the Python CALIFA $\textsc{starlight}$ 
Synthesis Organizer), producing the results discussed throughout this paper.
For the LArr data, the process was similar, 
and is described in detail in \citetalias{cortijo17}. 

The results reported in this paper rely 
on the GM and CB model bases 
described in \citetalias{cortijo17}. 
Basically, the GM base combines 
the $\textsc{granada}$ models of 
\cite{gonzalezdelgado2005} with \cite{vazdekis2010}, and is 
based on the Salpeter IMF. 
The CB base is built from an update of the \cite{bruzual&charlot2003} models, 
with a Chabrier IMF. 
Reddening is modelled as a foreground 
screen parametrized by A$_{V}$ (same reddening 
for all the SSP components) and following \cite{calzetti2000} 
reddening law with R$_{V}$ = 4.5, which was derived from observations of starburst galaxies, 
and which is found to be the most appropriate 
for interacting galaxies \cite{smith2016}. 

The quality of the spectral fits was examined using the $\overline{\Delta}$ 
indicator (\citealt{cidfernandes2013}).
We found median $\overline{\Delta}$ values of 6$\%$ and 5$\%$ for the CALIFA, 
and the LArr data, 
respectively. In the remainder of the paper, only the spaxels 
with $\overline{\Delta}$ $<$ 10$\%$ are shown in the maps, 
and considered in the analysis.

The effects of the Voronoi binning are present in all 
panels, but they are more important in the maps of 
luminosity-independent properties, such as the $A_{V}$ map.
This is because the maps of extensive properties  
(luminosity and mass) were ``dezonified"
by scaling the value at each $xy$ spaxel by its fractional 
contribution to the total flux in a zone (z index in equation 2). 
For example, for the stellar mass surface density we have applied:
\begin{equation}
\mu_{xy}=\frac{M_{xy}}{A_{xy}}=\frac{M_{z}}{A_{z}} \times w_{xyz}
\end{equation}
where $M_{xy}$ ($M_{z}$) is the stellar mass in a spaxel (zone), 
and $A_{xy}$ ($A_{z}$) denotes the area in that spaxel (zone), and
\begin{equation}
w_{xyz}=\frac{F_{xy}}{\sum _{xy|z} F_{xy}}
\end{equation}
where $F_{xy}$ is the mean flux at 
5635 $\pm$ 45 $\rm \AA$  (for more details 
see Section 5.2 of \citealt{cidfernandes2013}).
This operation was applied to luminosity and mass related
quantities, producing somewhat smoother images than 
obtained with $w_{xyz}=1$.
However, intensive properties such as $A_{V}$ and mean age 
cannot be dezonified.

\subsection{Results: Stellar mass and stellar mass surface density}\label{3.2}
An important $\textsc{starlight}$ output is the stellar 
mass, calculated from the stellar luminosity and 
taking into account the spatial variation of the mass-to-light ratio.
The total stellar mass obtained by summing the mass of each zone 
is $M_{\star} = \sum M_{z}$ = 5.4$\times$10$^{10}$ M$_{\odot}$.
This is the mass locked in stars at the observation epoch. 
Counting also the mass returned by stars to the interstellar 
medium, $M_{\star}'$ = 7.5$\times$10$^{10}$ M$_{\odot}$ 
were involved in star formation. These results are for the GM base 
(Salpeter IMF). With the CB base (Chabrier IMF), 
we find $M_{\star}'$ = 4.4$\times$10$^{10}$ M$_{\odot}$ and
$M_{\star}$ = 2.4$\times$10$^{10}$ M$_{\odot}$. 
\footnote{The ratio of the stellar masses derived with 
GM and CB models, $M_{\star} (GM)/M_{\star} (CB)$ = 2.25, is 
larger than the 1.78 value expected due to the different IMFs 
(Salpeter vs. Chabrier). This is because this ratio is also 
affected by the differences in the stellar population properties, 
specially the stellar ages. This is shown in Appendix B, where 
we discuss the uncertainties related to the models choice.}

We note that we are not underestimating the stellar masses with 
the optical spectral synthesis, since they 
are similar to the 
dynamical mass estimated by \cite{privon2013}, 
$M_{dyn} \lesssim$ 6$\times$10$^{10}$ M$_{\odot}$, 
which represents a lower limit as it is 
derived using a numerical model of the interaction.

Figure 4 shows the stellar mass surface 
density maps, $\mu_{\star}$ (in $M_{\odot} pc^{-2}$).
The upper left panel is the full CALIFA map. The white dashed 
lines indicate the positions of 1, 2, and 3 
Half Light Radius (HLR) \footnote{The HLR is defined 
as the length of the radial aperture which 
contains half of the total light of the galaxy at the 
rest-frame wavelength.}. In this case, 
1 HLR is equivalent to 2.8 kpc in physical distance, which 
is approximately in agreement with the half light radius measured by the CALIFA 
collaboration (10.3 arcsec = 4.0 kpc, \citealt{walcher2014}) 
through the growth curve 
analysis in elliptical apertures of the SDSS r-band image. 

The upper right panels in Fig. 4 are the LArr map, and a zoom of the CALIFA map 
in the same region covered by LArr data. 
The 2D maps of the stellar population properties were 
azimuthally averaged to allow their radial 
variations to be studied using PyCASSO. 
Radial apertures of 0.1 HLR in width were used to extract 
the radial profiles. Expressing radial 
distances in units of HLR allows the profiles of NGC 2623 
to be compared on a common metric with the control samples of Sbc and Sc galaxies.
The lower panel shows the profile of log $\mu_{\star}$
as a function of the radial distance
in HLR, in red for LArr and black for CALIFA. 
The uncertainties are shaded in light red, 
and grey, and represent the standard 
error of the mean, calculated as the standard deviation divided 
by the square root of the number of points (N) in each radial distance bin.
The differences in the uncertainties between the inner and outer 
regions are visually not strong, because the larger the radius, 
the larger  N. In terms of dispersion (=standard deviation) we find 
that, with respect to the NGC 2623 centre (N=8, CALIFA dataset), 
the dispersion is a factor 2.2 larger at 0.5 HLR (N=38), 
2.9 larger at 1 HLR (N=68), and 4 larger at 2 HLR (N=128). 
For the LArr data, the dispersion is a factor 2.7 larger at 
0.5 HLR (N=59) than in the centre (N=8).
For comparison, the grey lines are the profiles derived for Sbc (solid) 
and Sc (dashed) 
spiral galaxies from CALIFA (\citealt{gonzalezdelgado2015}, 
\citetalias{gonzalezdelgado2015} hereafter), with masses consistent with that of NGC 2623. 
We also include the profiles of the early-stage mergers 
IC 1623 W (blue), NGC 6090 NE (orange) and NGC 6090 SW (green).

We find a negative trend of $\mu_{\star}$ with distance
using the CALIFA map:  
from log $\mu_{\star}$ ($M_{\odot}$ pc$^{-2}$) $\sim$ 3.8 in the 
nucleus itself 
to $\sim$ 1.3 in the outer parts, at 3 HLR. 
The log $\mu_{\star}$ values derived from the LArr data 
are 0.1--0.2 dex larger than those
derived from the CALIFA data, 
but still within the dispersion.
We find only minor differences in the stellar mass surface 
densities and the other stellar population properties 
(see Sections \ref{3.3} and \ref{3.4}) 
derived from CALIFA and LArr datasets, always within 
the uncertainties; they could be due to the different 
spatial and spectral resolution, 
and also to the effect of 
the age-A$_{V}$ degeneracy, which is the most 
difficult to break in dusty star forming galaxies like NGC 2623.

The central stellar mass density of NGC 2623 
is comparable to Sbc galaxies in CALIFA sample.
We have compared the inner (from 0 to 1 HLR) and outer (from 1 to 2 HLR) 
gradients of the stellar mass surface density (as defined in equations 
6 and 7 of \citetalias{gonzalezdelgado2015}) in NGC 2623 and in 
the control spirals:
$\Delta_{in}$ log $\mu_{\star}$ = $-1.22$, $-1.15$, $-0.99$ for 
NGC 2623, Sbc, and Sc galaxies, respectively; and 
$\Delta_{out}$ log $\mu_{\star}$ = $-0.80$, $-0.64$, $-0.54$ for 
NGC 2623, Sbc, and Sc galaxies, respectively.
The stellar mass density profile of NGC 2623 
is a bit steeper, but comparable to Sbc galaxies in 
the CALIFA sample (which have a dispersion of $\sim$ 0.24 dex). 
\begin{figure*}  
\begin{center}
\includegraphics[width=0.61\textwidth]{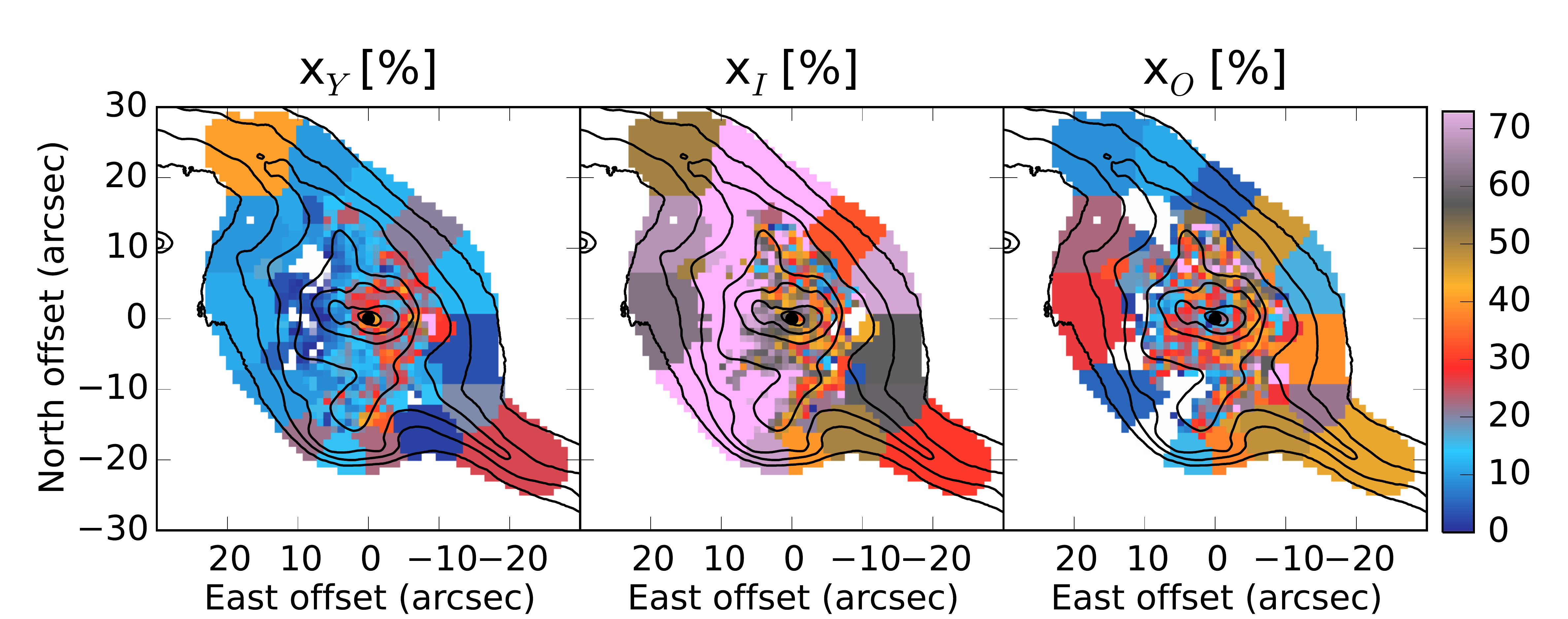} 
\includegraphics[width=0.38\textwidth]{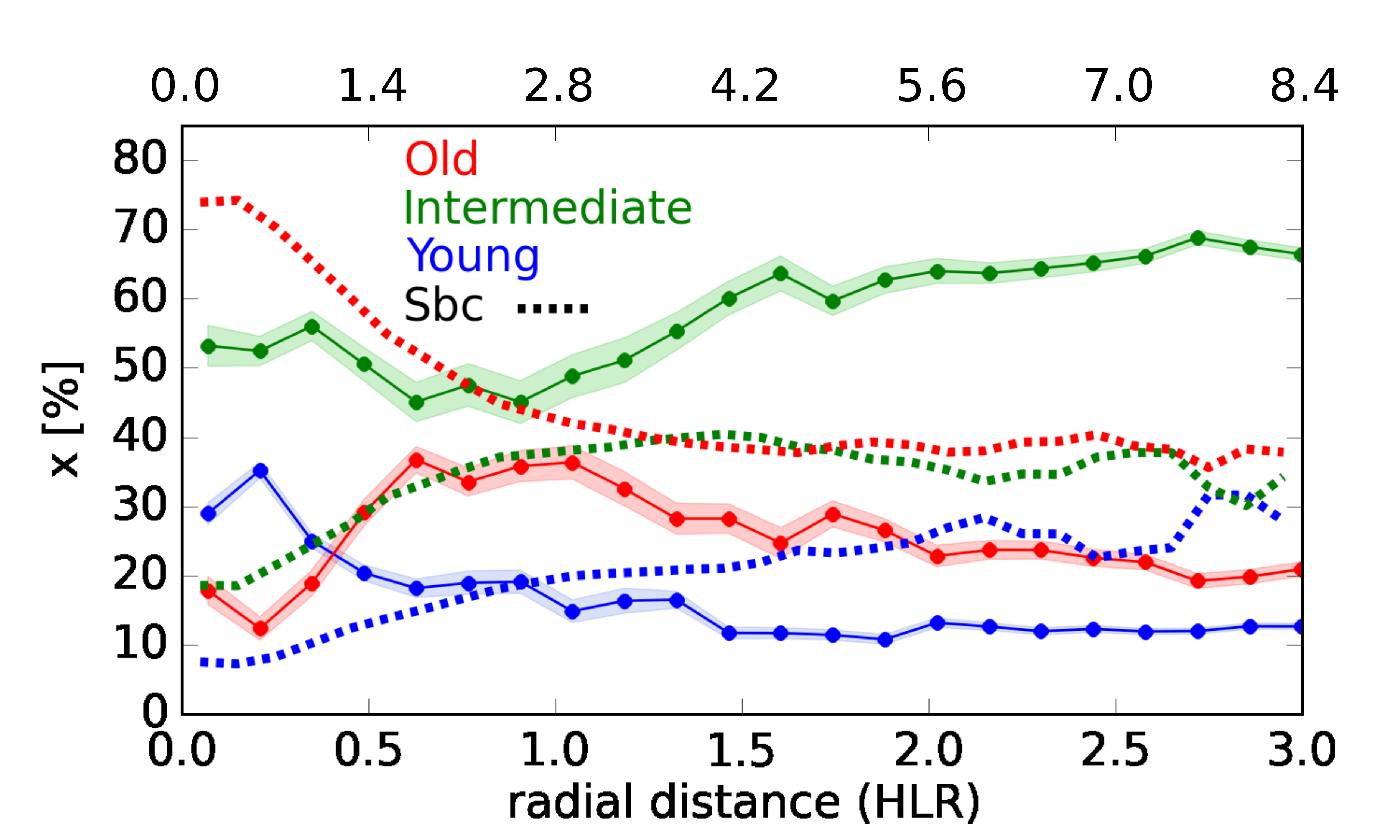}\\
\caption{Left panels: percentage contributions of young t $<$ 140 Myr , intermediate (140 Myr $<$ t $<$ 1.4 Gyr) 
and old  t $>$ 1.4 Gyr stellar populations to the observed light at 5635 $\rm \AA$.
Right panel: radial profiles of the light contributions from 
young x$_{Y}$ (blue), intermediate x$_{I}$ (green) 
and old populations x$_{O}$ (red) with the radial distance in HLR and
kpc. For comparison, the dashed lines with the same colours are the profiles 
for Sbc spiral galaxies from CALIFA.}
\label{Fig_4p16}  
\end{center}
\end{figure*} 
\begin{figure*}
\begin{center}
\includegraphics[width=0.61\textwidth]{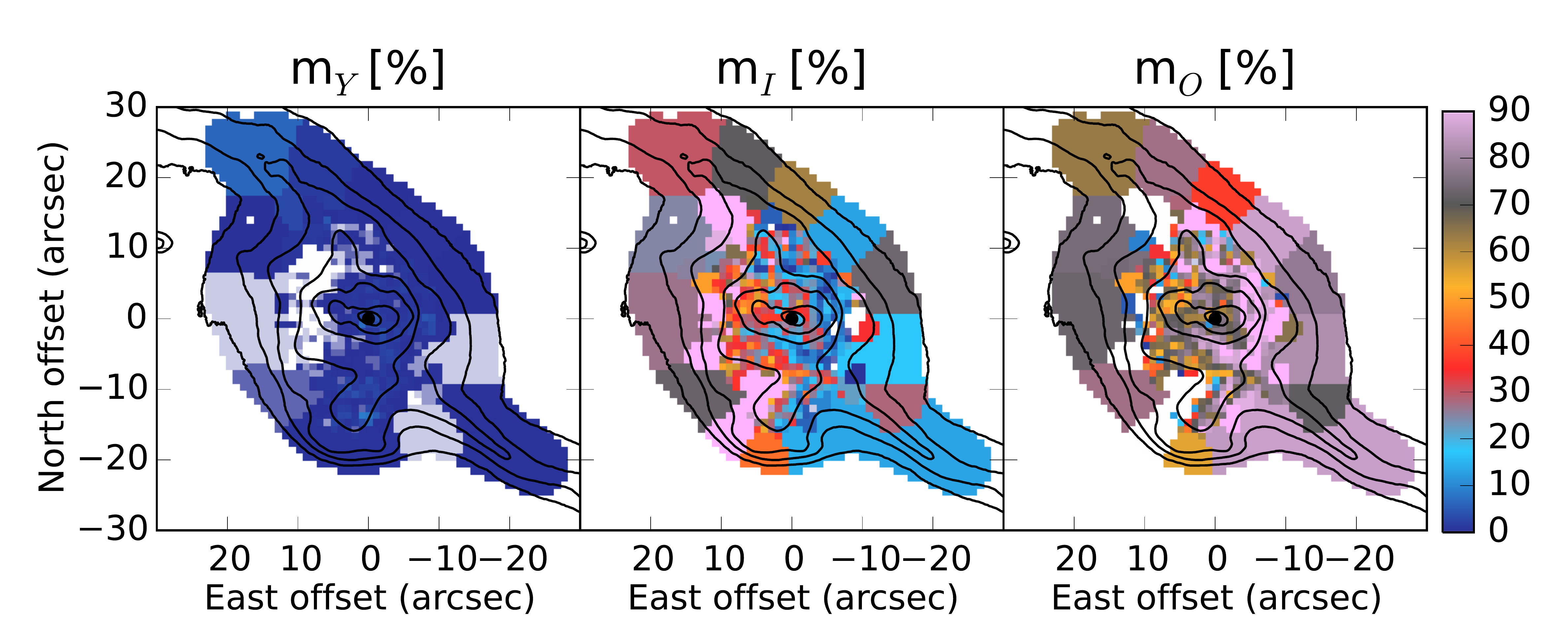} 
\includegraphics[width=0.38\textwidth]{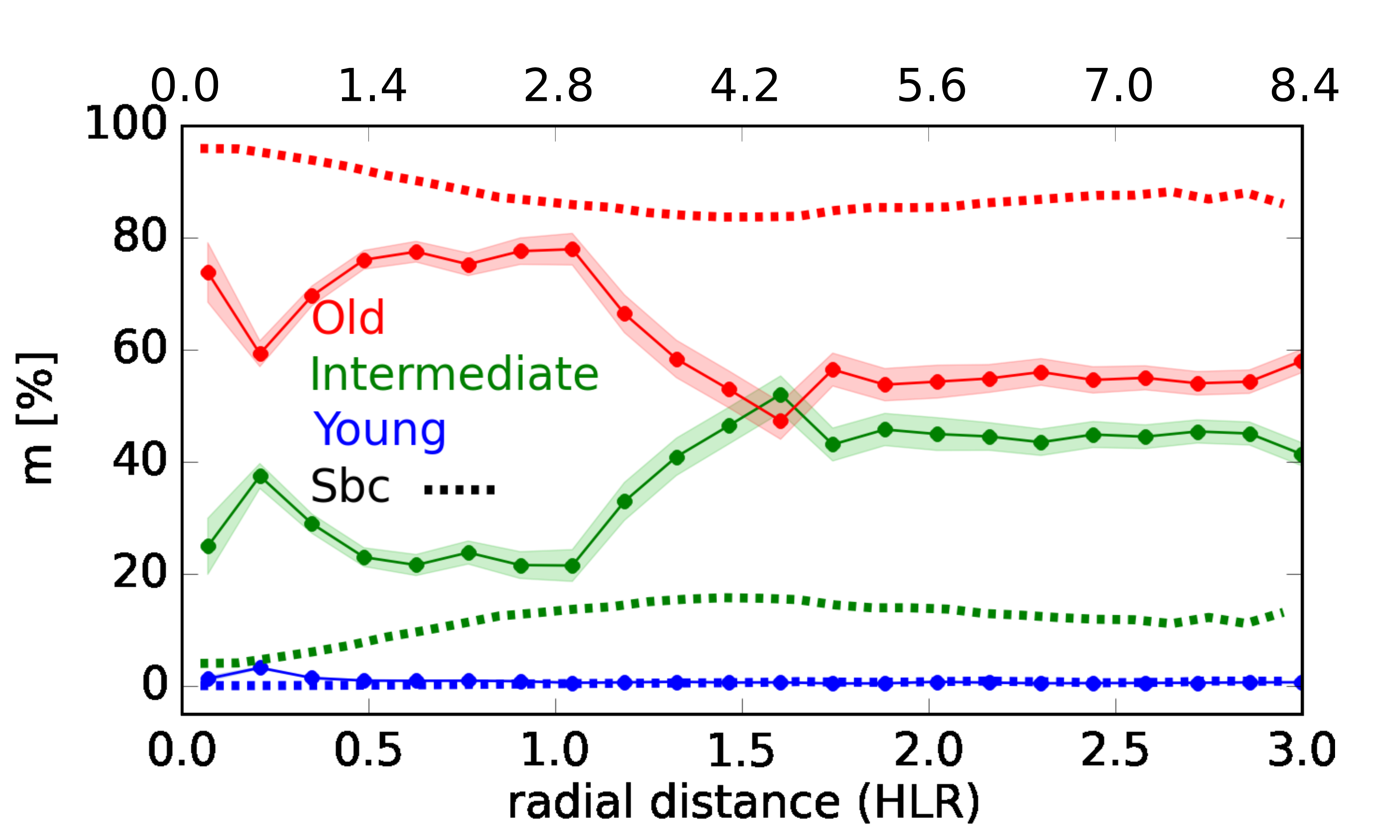}\\
\caption{As Figure 7 but for contributions to stellar mass.}
\label{Fig_4p18}  
\end{center}
\end{figure*}

\subsection{Results: Stellar dust extinction}\label{3.3}
$\textsc{starlight}$-based A$_{V}$ maps are shown in Figure 5, where 
the scale in CALIFA and LArr maps is the same 
to facilitate the comparison.

We find that CALIFA and LArr give similar results. 
For pixels at R = 0.5 $\pm$ 0.1 HLR,  we find 
$A_{V}^{LArr}$ = 1.38 $\pm$ 0.33 mag and 
$A_{V}^{PPAK}$ = 1.00 $\pm$ 0.23 mag, being both 
in agreement taking into account the dispersion.

NGC 2623 shows a negative gradient in A$_{V}$, as also 
observed in CALIFA Sb to Sc spirals. 
However, the absolute values, as well as the central gradient, 
are significantly larger in NGC 2623 than in spirals.

From the IFS data, we find that at 1 HLR, NGC 2623 stars are reddened by 
A$_{V}$ = 0.5 mag, which is comparable to 
Sbc--Sc galaxies considering the uncertainties.
The contrast is higher in the inner 0.2 HLR, where NGC 2623 is affected by 
A$_{V}$ = 1.4 mag, approximately 0.9 mag more than for Sbc--Sc galaxies. 
Moreover, the gradient in the inner HLR in Sbc--Sc galaxies 
is $\sim 0.3$ mag, while in NGC 2623 it is $\sim 0.9$ mag, three times 
larger.
We believe that this is due to the accumulation of gas 
and dust in the central regions, which occurs 
in the final stages of mergers.
This contrasts with the early-stage mergers 
IC 1623 and NGC 6090, 
where $A_{V}$ profiles are very flat, as expected if gas and dust 
are more uniformly distributed. Moreover, 
in both early-stage mergers, one of the progenitors 
is significantly more obscured than the other.
If LIRGs were originally normal late-type spirals, then 
we conclude that the merger process redistributes the gas and dust 
of these spirals such that most of it moves to one 
of the progenitors (probably the most massive). 
When the galaxies finally merge, most of the 
dust content is already concentrated in the central regions of the remnant. 

\subsection{Results: Ages}\label{3.4}
The mean light weighted log stellar 
age, $\langle \log \ age  \rangle_{L}$, 
was calculated using equation 3 
in \citetalias{cortijo17}.

Maps of $\langle \log \ age  \rangle_{L}$ are shown in Figure 6. 
From the CALIFA map we find that the youngest
regions are the region rich in star clusters
located south of the nucleus, and the northern tidal 
tail, with mean ages around $\sim$ 300 Myr. 
The nuclear region is $\sim$ 500 Myr old, and 
the south tidal
tail, and the eastern region of the main body, are 
older, with ages $\sim$ 1 Gyr.
In the circum-nuclear regions we find a positive 
trend of age with distance, from $\sim$ 500 Myr 
in the nucleus itself to $\sim$ 900 Myr at 1 HLR.
It is also interesting to note that the full range of
mean stellar ages is 
140 Myr to less than 2 Gyr.

Again, when comparing the two datasets (LArr and PPaK), we find that 
they are indistinguishable from each 
other, considering the dispersion.
At 0.5 $\pm$ 0.1 HLR,  we find 
$\langle \log \ age \rangle_{L}^{LArr} (yr)$ = 8.75 $\pm$ 0.28 and 
$\langle \log \ age \rangle_{L}^{PPAK} (yr)$ = 8.85 $\pm$ 0.12. 

The average light weighted age of NGC 2623 
at $\sim$ 1 HLR is $\sim$ 900 Myr, similar to the Sbc-Sc control 
galaxies.  
However, the nuclear region (within 0.2 HLR) of NGC 2623 
is much younger, 500 Myr old, compared with the 
3.6 Gyr (1.4 Gyr) old of the Sbc (Sc) control galaxies. 
Above 1 HLR the mean stellar ages of NGC 2623 are 
similar to those in Sbc--Sc galaxies.
The inner age gradient of NGC 2623 is positive with distance, 
in contrast with the negative gradient of Sbc--Sc galaxies:
$\Delta_{in} age$ $\sim$ 400 Myr, $-2.4$ Gyr, $-670$ Myr 
in NGC 2623, Sbc, and Sc galaxies, respectively.
The early-stage mergers IC 1623 and NGC 6090 are, 
on average, younger 
than Sbc--Sc galaxies and NGC 2623. Moreover, their age 
profiles are significantly flatter than in Sbc--Sc galaxies, but 
not inverted as in NGC 2623. 
 
If LIRGs were originally late-type spirals, then 
we conclude that the merger-induced star formation 
leads to a general rejuvenation of the 
progenitor galaxies 
during the early-stage merger stage, that affects not only the central
regions but also the ``disk''. 
When the galaxies finally merge, most of the gas is already 
in the central regions and the young star formation is therefore 
concentrated there, whereas the outer parts are evolving passively. 
This would be consistent with the positive age gradient in NGC 2623. 
\begin{figure*}
\begin{center}
\includegraphics[width=1.0\textwidth]{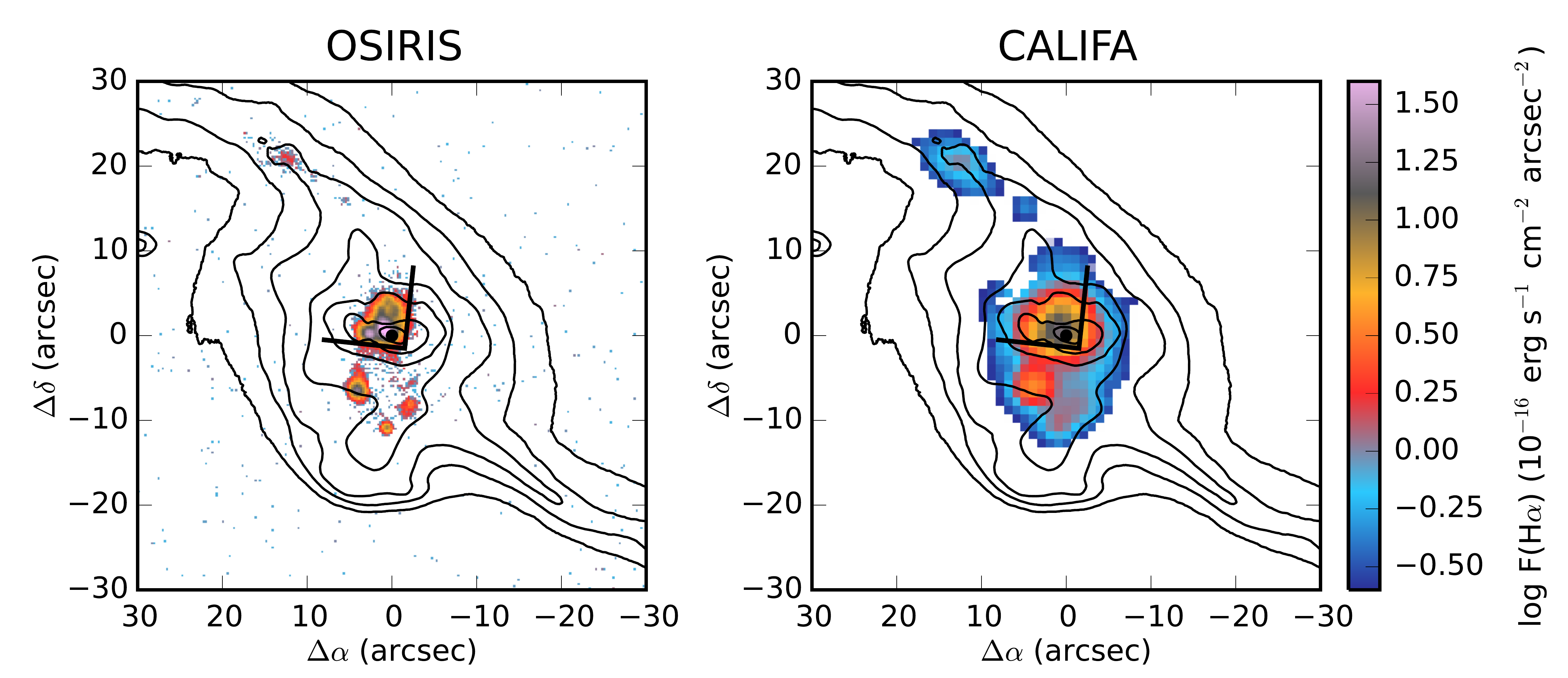}
\caption{OSIRIS H$\alpha$ (left) and CALIFA H$\alpha$ (right) emission line fluxes 
in units of 10$^{-16}$ erg s$^{-1}$ cm$^{-2}$ arcsec$^{-2}$ presented
on a logarithmic scale. 
The black cone delimits the nuclear 
outflowing nebula reported by \protect\cite{lipari2004}. 
The HST F555W image is shown in contours, smoothed to approximately 
match the spatial resolution of our IFS data.}
\label{Fig_4p22}  
\end{center}
\end{figure*}
\begin{figure}
\begin{center}
\includegraphics[width=0.5\textwidth]{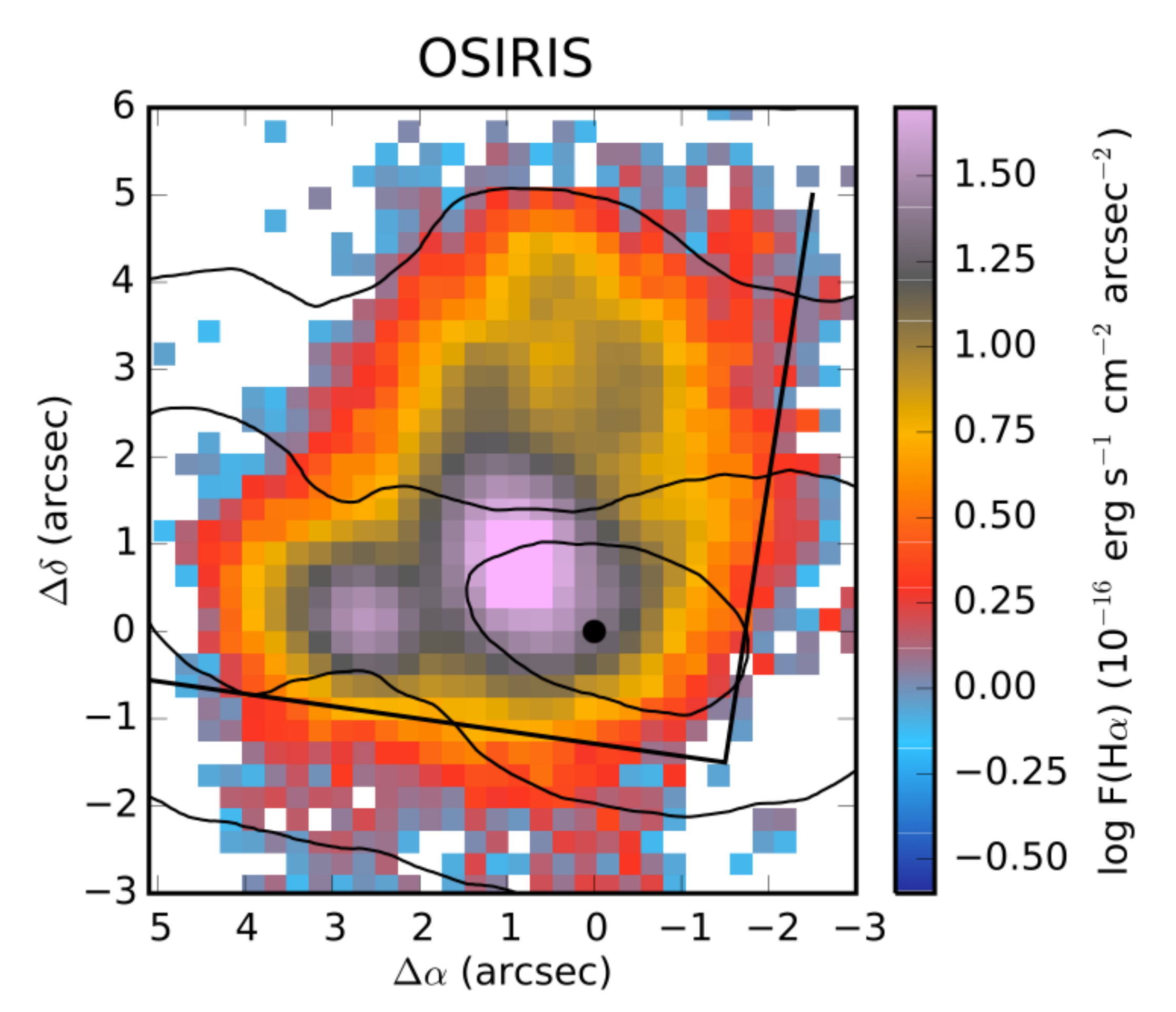}
\caption{Zoom of the OSIRIS H$\alpha$ emission line image for the nuclear region of NGC 2623. 
HST continuum contours are overplotted in black. 
The black dot marks the continuum centre.}
\label{Fig_4p23}  
\end{center}
\end{figure}

\subsection{Results: Contribution of young, intermediate and old
populations}\label{3.5}
In this section we study the spatially resolved 
SFH, by condensing the age distribution encoded in the population 
vector into three age ranges: young (YSP, t $\le$ 140 Myr), 
intermediate age (ISP, 140 Myr $<$ t $\le$ 1.4 Gyr), and 
old stellar populations (OSP, t $>$ 1.4 Gyr), as in 
\citetalias{cortijo17}. 

Figure 7 presents the maps (left) and radial profiles (right) 
of the light fractions at 5635 $\AA$ (the normalization $\lambda$) due 
to YSP, ISP, and OSP (x$_{Y}$, x$_{I}$, and x$_{O}$). 
In the right panels the blue, green, and red dashed lines are the 
contributions to light of YSP, ISP, and OSP of Sbc galaxies 
from the control sample.
The contribution to light of the YSP is specially important 
in the center, the north 
tidal tail and in some places in the region rich in star 
clusters located south of the nucleus, with contributions 
to total light of $\sim$ 40$\%$.
The ISP dominate the light almost everywhere 
(with contribution up to 70$\%$).
It is interesting to note the presence of this widespread 
ISP in this merger. The OSP are important to the southeast 
of the system, with contributions to light of $\sim$ 40--50$\%$. 
From the radial profiles we find that the 
light is everywhere (0--3 HLR) dominated by ISP, 
with contributions of 50--70$\%$, 
in contrast to the 20--40$\%$ in Sbc galaxies. 
We also find that within 0.5 HLR 
there exist a significant contribution to light by YSP, 
of $\sim$ 20--30$\%$, in contrast to the $<$ 10$\%$ YSP 
contribution in Sbc galaxies.

Similarly, in Figure 8 we present the maps (left) and radial 
profiles (right) 
of the mass fractions due to YSP, ISP, and 
OSP (m$_{Y}$, m$_{I}$ and m$_{O}$). 
The fraction of stellar mass contributed 
by YSP is less that 5$\%$ everywhere.
Most of the mass is contained in OSP.
In particular, OSP dominate the mass contributions within 
1.5 HLR (50--80$\%$), significantly smaller than 
the 84--96$\%$ contribution of OSP in Sbc galaxies.
Beyond 1.5 HLR, the mass contribution of the ISP 
is also important, between 40--50$\%$, 
and significantly higher than the 15$\%$
contribution of ISP to the mass in Sbc galaxies.

We have also computed the total mass in YSP, ISP, and OSP. 
Of the total stellar mass in NGC 2623 
($5.4 \times10^{10}$ M$_{\odot}$), 
the mass in young, intermediate age, and old 
populations is $6.6 \times 10^{8}$ ($1\%$), 
$1.2 \times 10 ^{10}$ ($22\%$), and 
$4.1 \times 10 ^{10}$ M$_{\odot}$ ($77\%$), respectively.

The mass in young components (of $\lesssim$ 300 Myr) derived 
some the IFS data is comparable to the approximate mass in 
star clusters derived from the photometry, 
$M_{clus}^{NGC 2623}$ $\sim$ 6 $\times$ 10$^{8}$ M$_{\odot}$ 
(see Appendix A). 
This supports the hypothesis that the vast majority 
of stars form in clusters rather than in isolation \citep{chandar2015}.
\begin{figure}
\begin{center}
\includegraphics[width=0.5\textwidth]{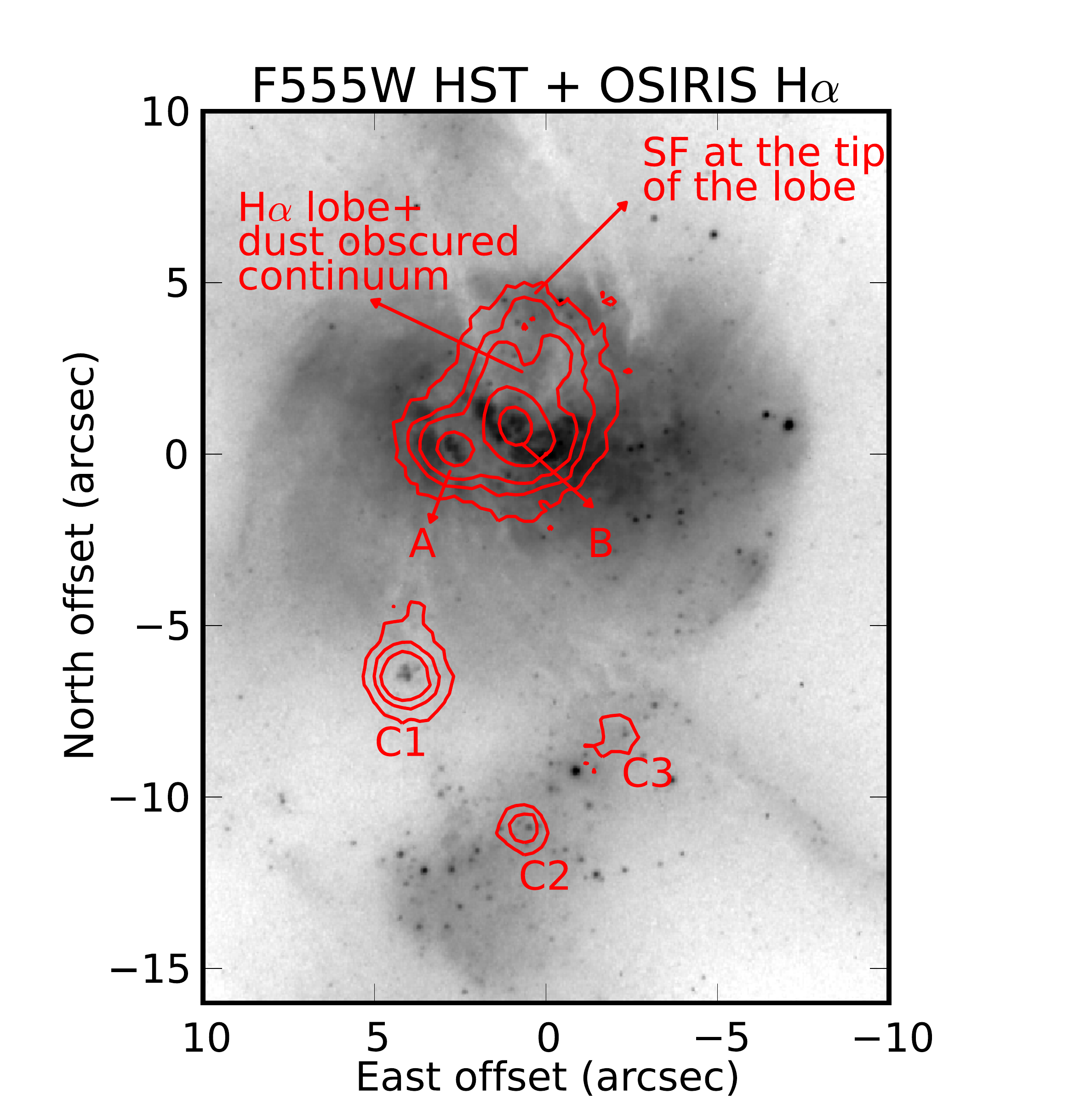}
\caption{The HST F555W continuum image of the nuclear region of NGC 2623 together 
with the OSIRIS H$\alpha$ line emission plotted as red contours. 
The star clusters aggregations A and B, and the HII regions C1, C2, and C3 are labelled. 
The positions of the H$\alpha$ lobe and star formation (SF) associated to the tip of the lobe 
are also marked.}
\label{Fig_4p24}  
\end{center}
\end{figure}
\begin{figure}
\begin{center}
\includegraphics[width=0.5\textwidth]{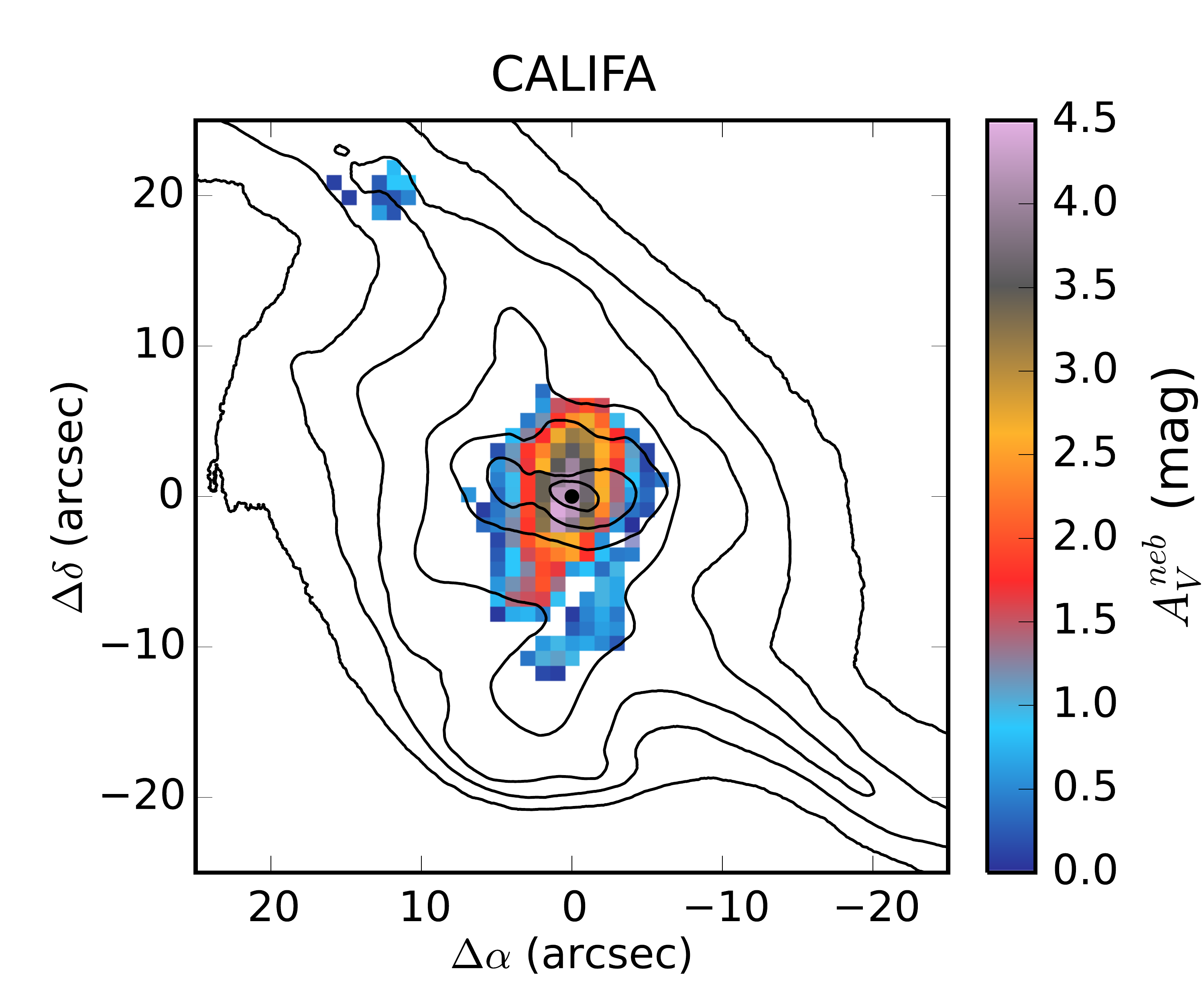}
\caption{$A_{V}^{neb}$ map derived from the H$\alpha$/H$\beta$ emission line ratio assuming a 
\protect\cite{calzetti2000} dust attenuation curve.}
\label{Fig_4p28}  
\end{center}
\end{figure}

\section{Ionized gas emission}\label{4}

In this section we focus on the ionized 
gas properties. In order to account for stellar 
absorption, and to obtain pure nebular emission 
line spectra we subtracted {\sc starlight} best fits 
from the spectra. 
From the spectra we measured the line fluxes of the
most prominent 
emission lines ([OII]$\lambda$3727, H$\beta$, 
[OIII]$\lambda$5007, [OI]$\lambda$6300, H$\alpha$, 
[NII]$\lambda$6584 and [SII]$\lambda$$\lambda$6716,6731), 
by fitting them with Gaussian profiles. 
Emission-line intensity maps were thus created for 
each individual line. Only spaxels with signal to noise ratio 
(SNR) $>$ 3 are considered in the following analysis.
Hereafter, we focus only on the CALIFA IFS data, and 
on the higher spatial resolution H$\alpha$ and 
[NII]$\lambda$6584 emission line maps from the OSIRIS TF 
(resolution $\sim$ 1'' compared with 2.4'' for CALIFA).

\subsection{Ionized gas morphology}\label{4.1}
Figure 9 shows the  H$\alpha$ maps derived  from the
OSIRIS (left panel) and CALIFA (right panel) data. 
These trace the distribution of ionised gas.
The emission line distribution in the OSIRIS map is more compact 
than that in the CALIFA map, mainly due to its better 
spatial resolution. Moreover, we note that the CALIFA H$\alpha$ 
emission has been corrected for the underlying stellar absorption,
while this was not possible in OSIRIS data. We estimate 
that $\sim$15$\%$ of the H$\alpha$ flux is missing 
in OSIRIS data due to this.
In NGC 2623 the vast majority of the ionized gas 
emission comes from the nuclear regions. 
Outside the nuclear regions, we detect three 
star forming clusters to the south, 
in the cluster-rich region found by \cite{evans2008}. 
The latter represents the youngest of the 
$\sim$ 100 clusters in this region. Fainter H$\alpha$ 
emission is also detected in the northern tidal tail, 
in both CALIFA and OSIRIS maps. 

L\'ipari et al. (2004, hereafter \citetalias{lipari2004}) reported 
convincing evidence of the existence 
of an outflow due to a nuclear dusty starburst in this system, 
leaning on the broad (FWHM $\sim$ 600 km s$^{-1}$), 
blueshifted ($\sim$-400 km s$^{-1}$) and 
highly asymmetric line profiles at extranuclear 
locations (see their Fig.7, and Fig. 11 bottom). The 
enhanced [NII]/H$\alpha$ ratio, is also consistent with a scenario 
in which a different excitation mechanism, most naturally shocks, 
dominate at the location of the outflowing gas.
The orientation and location of the 
outflow, as reported by \citetalias{lipari2004}, are indicated with black solid 
lines in Figure 9. It forms a cone-shaped nebula, with an opening angle 
of $\theta \sim$ 100$\degree$.

With respect to the morphology, and given the higher spatial 
resolution of the OSIRIS map, we are able to 
distinguish substructures within the cone-nebula 
that are unresolved in the CALIFA and \citetalias{lipari2004} maps. 
Figure 10 zooms in on the nuclear regions of the OSIRIS 
H$\alpha$ map. 
Also, to get a better idea of the overall 
morphology of the cone nebula, in Figure 11 we have superimposed 
in red contours the OSIRIS H$\alpha$ emission over 
the HST F555W continuum image. 
Within the cone nebula, we notice different structures.
There is a point-like source offset with respect to the 
continuum centre by (2.7$\tt{''}$E, 0.1$\tt{''}$N, label A in Figure 11). 
Its morphology is consistent with a star forming knot, also 
confirmed by the line-ratios ([NII]/H$\alpha$ $\sim$ 0.5, and 
[OIII]/H$\beta$ $\sim$ 1.25).
There is also an extended lobe-like emission region with 
bow-shaped, clumpy ends that seems to emerge from the nucleus. 
This kind of morphology is typical of shocked regions.
Inside this lobe, we find that the brightest H$\alpha$ emission 
corresponds to several cluster aggregations, labelled B.
Most of the extended H$\alpha$ emission is spatially 
coincident with highly obscured regions. Some bright 
clusters appear at the tip of the lobe (at the 
``edges" of the outflow), whose formation 
might have been induced by the outflow.
Finally, the three HII regions south of the nucleus 
(C1, C2, and C3) are also indicated in the Figure, and 
their stellar counterparts are clearly detected.

\subsection{[SII]-based gas densities}\label{4.2}
We have estimated the electron density of the 
emitting gas ($n_{e}$) in the centre of NGC 2623, 
using the ratio [SII]$\lambda$6716/[SII]$\lambda$6731, and 
applying \cite{osterbrock2006} calibration assuming 
an electron temperature of T = 10$^{4}$ K.
On average, the ratio is around the low 
density limit ($\sim$ 1.4 $\pm$ 0.1), indicating  densities  
of $\sim$100 cm$^{-3}$ or lower. 
The ratio has a value around 1.25 $\pm$ 0.08 in the nucleus itself, 
equivalent to an electron density 189 $\pm$ 97 cm$^{-3}$. 
\begin{figure*}
\begin{center}
\includegraphics[width=0.9\textwidth]{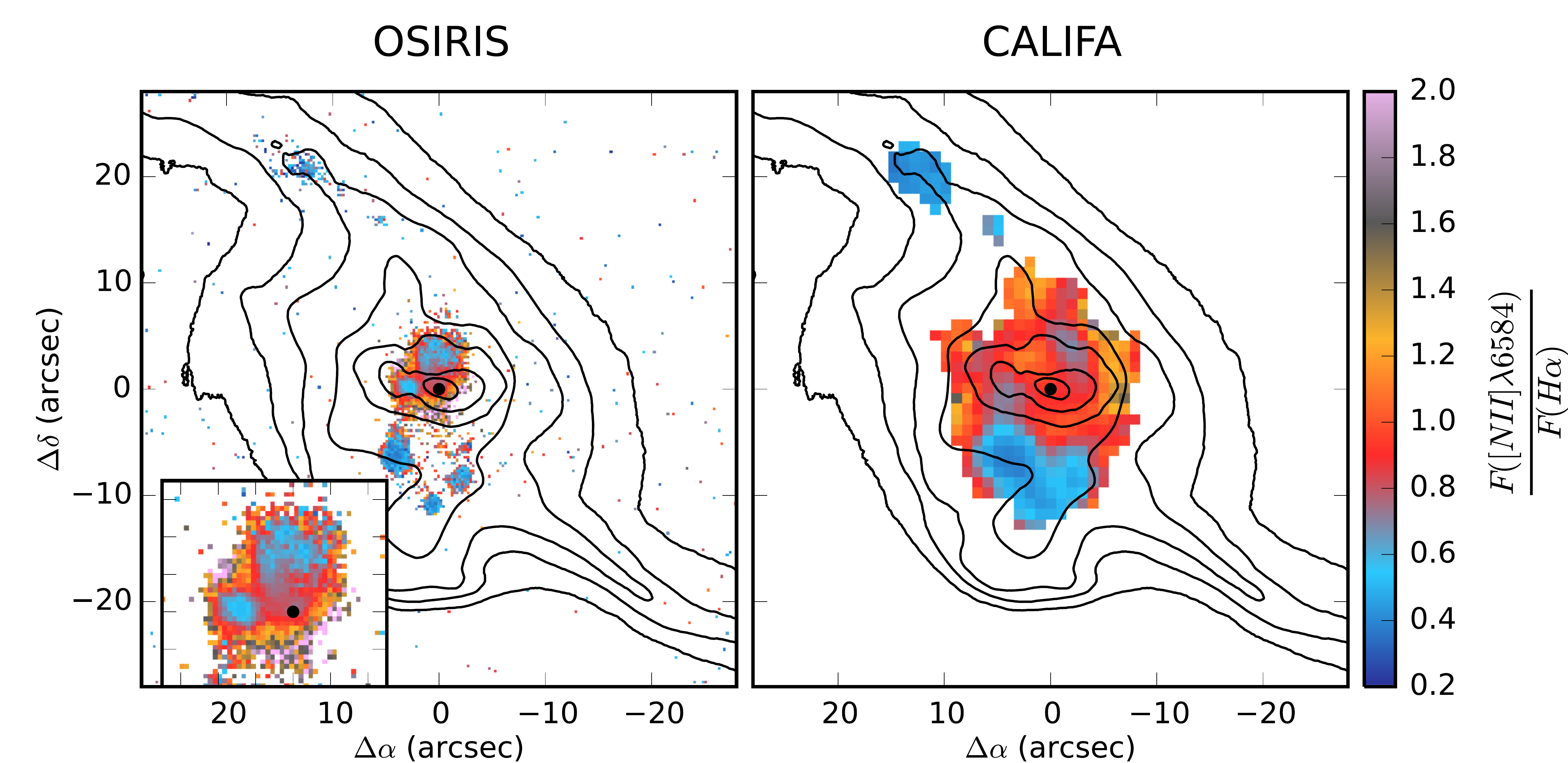}
\caption{Left: emission line ratio map of [NII]$\lambda$6584/H$\alpha$ derived from the OSIRIS data. 
The box in the lower left corner is a zoom of the nuclear region. 
Right: [NII]$\lambda$6584/H$\alpha$ derived from the CALIFA data.}
\label{Fig_4p28}  
\end{center}
\end{figure*}
\begin{figure*}
\begin{center}
\includegraphics[width=0.9\textwidth]{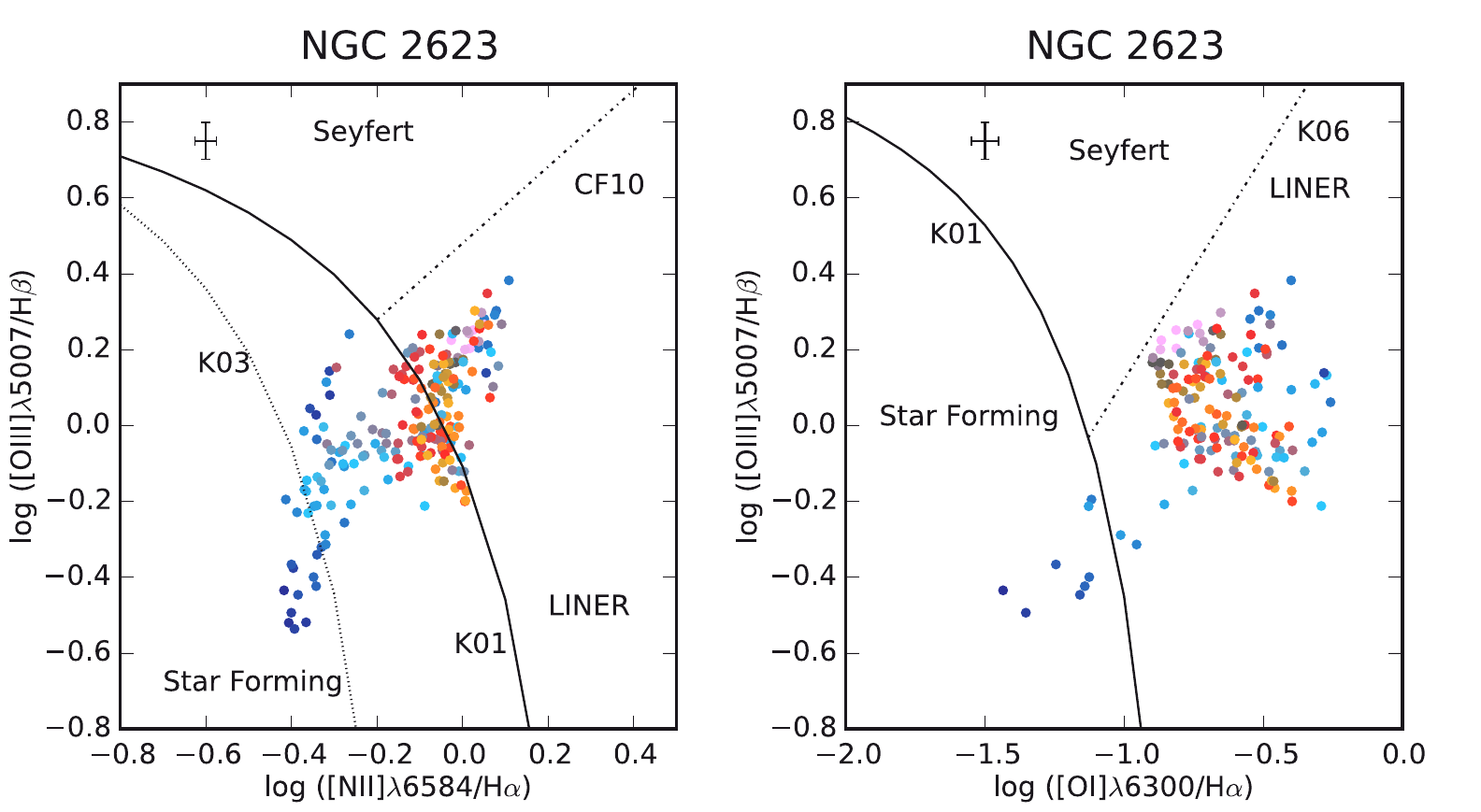}
\caption{[NII]$\lambda$6584/H$\alpha$ vs.\ 
[OIII]$\lambda$5007/H$\beta$ (left panel), and 
[OI]$\lambda$6300/H$\alpha$ vs.\
[OIII]$\lambda$5007/H$\beta$ (right panel) line ratio diagnostic 
diagrams. Spaxels are colour coded according to the H$\alpha$ velocity 
dispersion map shown in Fig. 15.
Solid lines in both diagnostic diagrams mark 
the ``maximum starburst lines" obtained by \protect\cite{kewley2001} (K01) 
by means of photoionization models of synthetic starbursts, the 
dotted line in the left panels marks the pure SDSS star-forming galaxies 
frontier semi-empirically 
drawn by \protect\cite{kauffmann2003} (K03), and the dashed-dotted lines 
that separate Seyferts from LINERs are 
from \protect\cite{cidfernandes2010b} (CF10) 
and \protect\cite{kewley2006} (K06).
The average error bars are shown in the top left corners.}
\label{Fig_4p31}  
\end{center}
\end{figure*}

\subsection{Ionized gas dust attenuation}\label{4.3}
We estimate the nebular extinction, $A_{V}^{neb}$, 
assuming an intrinsic ratio of I(H$\alpha$)/I(H$\beta$)=2.86, 
valid for Case B recombination for an electron temperature T = 10000 K and an
electron 
density 10$^{2}$ cm$^{-3}$ \citep{osterbrock2006}, 
typical of star forming regions. 
We used the \cite{calzetti2000} reddening function, 
which works better for starburst galaxies.

In Figure 12 we show the $A_{V}^{neb}$ map.
The highest extinction is found in the nuclear
regions, where $A_{V}^{neb}$ is up to 4.5 magnitudes.
This level of extinction implies that a dust 
screen with high covering fraction lies in front 
of the central region of the galaxy, consistent with the 
dust lanes visible in the HST images.
However, the ionized gas in the northern tidal tail and
in the clusters south of the nucleus has
$A_{V}^{neb} \le$ 1 mag. 
The global average is $A_{V}^{neb}$ $\sim$ 1.4 mag.

We have compared the ratio of the nebular 
and the stellar extinction, using 
the spaxels where the SNR of H$\alpha$ and H$\beta$ 
is above 3, and $A_{V}^{stars}$ > 0.1 mag, 
and find an average of
$\langle \frac{A_{V}^{neb}}{A_{V}^{stars}} \rangle = 1.7 \pm 1.2$. 
This result is in agreement with what we found 
in the early-stage merger 
LIRGs IC 1623 W and NGC 6090, and with other 
results from the literature \citep{calzetti1994,kreckel2013}.

\subsection{Emission line ratios sensitive to ionisation source}\label{4.4}
Standard line ratios sensitive to the shape of 
the ionising spectrum have been calculated. 

First, we note that the [OIII]$\lambda$5007/H$\beta$ ratio 
is below $\sim$2 everywhere. Higher values would be expected for 
Seyfert-like emission \citep{heckman1987,veilleux1995}. 
We point out that in our spatially 
resolved observations, the emission 
lines do not show Seyfert-like line ratios, not even 
in the nuclear region (see also next Section).
Although X-ray and IR data point to the existence of 
an AGN embedded in dust 
within the inner kiloparsec of the system, the bulk of infrared 
(and thus bolometric) luminosity is generated by star 
formation \citep{evans2008,petric2011}.

In Figure 13 we show the [NII]$\lambda$6584/H$\alpha$ 
maps for OSIRIS data (left) and CALIFA data (right).
The lowest values of the ratio are found in the north 
tidal tail and in the three star forming knots south of the 
nucleus (C1, C2, and C3). 
The CALIFA ratios are consistent with those derived from the 
OSIRIS data, but less sensitive 
to small-scale variations due to the coarser spatial resolution. 
With the greater spatial resolution of OSIRIS, we see 
finer details of the ionization in the nuclear region of NGC 2623.
In particular, inside the \citetalias{lipari2004} cone nebula, the nuclear knot (A), 
and some pixels at the clumpy ends 
of the nuclear lobe, are also consistent with 
star formation ionization ([NII]/H$\alpha$ $\sim$ 0.5), 
while the remainder have [NII]/H$\alpha$ 
between 1.0 and 1.2.
For the spaxels at the edges of the cone nebula, 
the [NII]/H$\alpha$ ratio  is even larger, with
values in the range 1--2, which could be associated with the 
effects of shocks produced by the outflow
colliding with the ISM \citep{heckman1987,heckman1990}.

We note that our measurements are in
perfect agreement with those reported by \citetalias{lipari2004} 
inside the cone nebula. However, outside the cone nebula, 
\citetalias{lipari2004} report values of [NII]/H$\alpha$ $>$ 3, 
which is not consistent with our results. 
As previously mentioned, in \citetalias{lipari2004} the H$\alpha$ flux 
is not corrected for the underlying 
stellar absorption, and therefore, 
the H$\alpha$ flux is underestimated, 
leading to higher values of this ratio. This effect 
is stronger outside the cone nebula, where the 
H$\alpha$ emission is weaker. From the IFS 
we calculate that 20--25$\%$ of the H$\alpha$ flux 
is underestimated when not correcting from the stellar 
absorption, and the morphology is more compact and more similar 
to \citetalias{lipari2004} and OSIRIS map in that case. We note how appropriate is 
IFS for this kind of analysis.

The [OI]$\lambda$6300/H$\alpha$ and 
[SII]$\lambda \lambda$6716+6731/H$\alpha$ maps 
inside the cone nebula have values $\sim$ 0.2 and $\sim$ 0.5, respectively, 
consistent with \citetalias{lipari2004}.
Following \cite{heckman1990}, these
values are indicative of shock heated gas of 
moderate velocity. These results are confirmed by 
the diagnostic diagrams shown in next Section.

\subsection{Diagnostic diagrams}\label{4.5}
We have used diagnostic diagrams to 
identify the nature of the dominant ionization mechanisms. 
Fig. 14 presents two diagrams: 
[NII]$\lambda$6584/H$\alpha$ vs.\  [OIII]$\lambda$5007/H$\beta$ 
(left panel, the ``BPT diagram'', \citealt{baldwin1981}), and 
[OI]$\lambda$6300/H$\alpha$ vs.\  [OIII]$\lambda$5007/H$\beta$ 
(right panel, \citealt{veilleux&osterbrock1987}). 
The spaxels are colored according to their 
H$\alpha$ velocity 
dispersion ($\sigma_{H\alpha}$), following the scale shown 
in the $\sigma_{H\alpha}$ map in Fig. 15 (top panel).

Overplotted as black lines are empirically and
theoretically derived separations between LINERs/Seyferts and 
HII regions.
The dotted line is from \cite{kauffmann2003}, 
a conservative demarcation to separate SDSS star-forming galaxies 
from those hosting an AGN; the solid line is from \cite{kewley2001},
marking the ``maximum starburst lines'' derived using the results of 
photoionization models of synthetic starbursts.
Finally, the dashed-dotted lines that separate Seyferts from 
LINERs are from \cite{cidfernandes2010b} and \cite{kewley2006}.

Only the northern tidal tail and 
the brightest knot south of the nucleus have line ratios consistent with
pure stellar photoionization.
On the other hand, the gas inside the nuclear cone defined by \citetalias{lipari2004} 
has LINER-like emission lines. This is consistent 
with the presence of the outflow reported by \citetalias{lipari2004}. 
In general, the higher the velocity dispersion, the more 
LINER-like the ionization is, suggesting shocks as the 
cause, as previously found in other 
local U/LIRGs \citep{monreal-ibero2006,rich2014}, 
and in the two LIRGs in \citetalias{cortijo17}. 
An ensemble of HII regions would not be able to reproduce 
the line ratios we observe. However, many spaxels fall 
in the composite region of the BPT, and according to the highest resolution 
OSIRIS and HST data, a collection of HII regions is also present in the 
plane of the galaxy. Thus, the presence of both an outflow and 
HII regions is necessary to explain the morphology and 
mixture of ionization mechanisms.
In addition, we note that there is also a region to the west 
of the nucleus with line ratios consistent with LINER-like 
ionization, although this region has a low velocity dispersion, 
more typical of HII regions.

\subsection{Gas kinematics}\label{4.6}
We have determined the kinematics of the ionized gas in NGC 2623 
using the Gaussian fits to H$\alpha$.
To obtain the final velocity dispersion, we have subtracted the 
CALIFA instrumental resolution ($\sigma_{inst}$) in quadrature, 
where $\sigma_{inst}$ at H$\alpha$ is $\sim$ 116 km s$^{-1}$.
Therefore, line widths below $\sim$ 58 km/s are 
unresolved at the CALIFA resolution. 

The H$\alpha$ velocity dispersion map is shown in the top panel 
of Figure 15. $\sigma_{H\alpha}$ is low ($<$ 100 km s$^{-1}$) 
in the star forming regions to the south of the nucleus, 
and in the northern tidal tail. The highest values of the velocity 
dispersion are reached inside the cone nebula 
defined by \citetalias{lipari2004} (up to 220 km/s). 
This high sigma could be a simple consequence of the two velocity 
components in H$\alpha$ found by \citetalias{lipari2004}, that clearly 
prove the presence 
of an outflowing gas in NGC 2623. In our data, which have worse spectral 
resolution, these two components are convolved into an unique 
H$\alpha$ line, that is significantly broader than in the surrounding 
ionized gas \footnote{We have analysed in more detail the H$\alpha$+[NII] 
line profiles in the outflow region and found that they are better fit 
when including a second broad (FWHM $\sim$ 670 km/s) component, 
blueshifted by $\sim$-200 km/s with 
respect to the narrow systemic component. 
The two components fit left less residuals that the one component fit, 
in agreement with \citetalias{lipari2004} findings.}.  
Obviously, a fraction of the broadening may also be due to unresolved 
non-circular motions in the disk and/or to unresolved central rotation 
due to beam smearing. However, since  the data of \citetalias{lipari2004} prove 
so clearly the existence of an outflow, we are more tented to 
interpret our sigma map as a consequence of this outflow.

In the bottom panel of Fig. 15 we show the H$\alpha$ velocity field. 
A rotation pattern is preserved in 
the nucleus, with an amplitude of $\pm$ 120 km/s, in 
agreement with \cite{barrera-ballesteros2015}. 
We have found that the kinematic center is displaced 2 arcsec 
to the east with respect to the photometric center. However, 
this is within our spatial resolution. 
The position angles of the kinematic major axis for the 
approaching a receding sides calculated by 
\cite{barrera-ballesteros2015} are shown in the 
Figure with black solid lines. For NGC 2623, 
PA$_{kin}^{appro}$ = 83.1$\degree$ $\pm$ 5.9$\degree$ and 
PA$_{kin}^{reced}$ = 115.1$\degree$ $\pm$ 11.4$\degree$.
Beyond the nucleus, both the star forming regions in the 
north tidal tail 
and the material in the cluster-rich region 
to the south are approaching to us at $\sim$ 40 km/s.

\begin{figure}
\begin{center}
\includegraphics[width=0.5\textwidth]{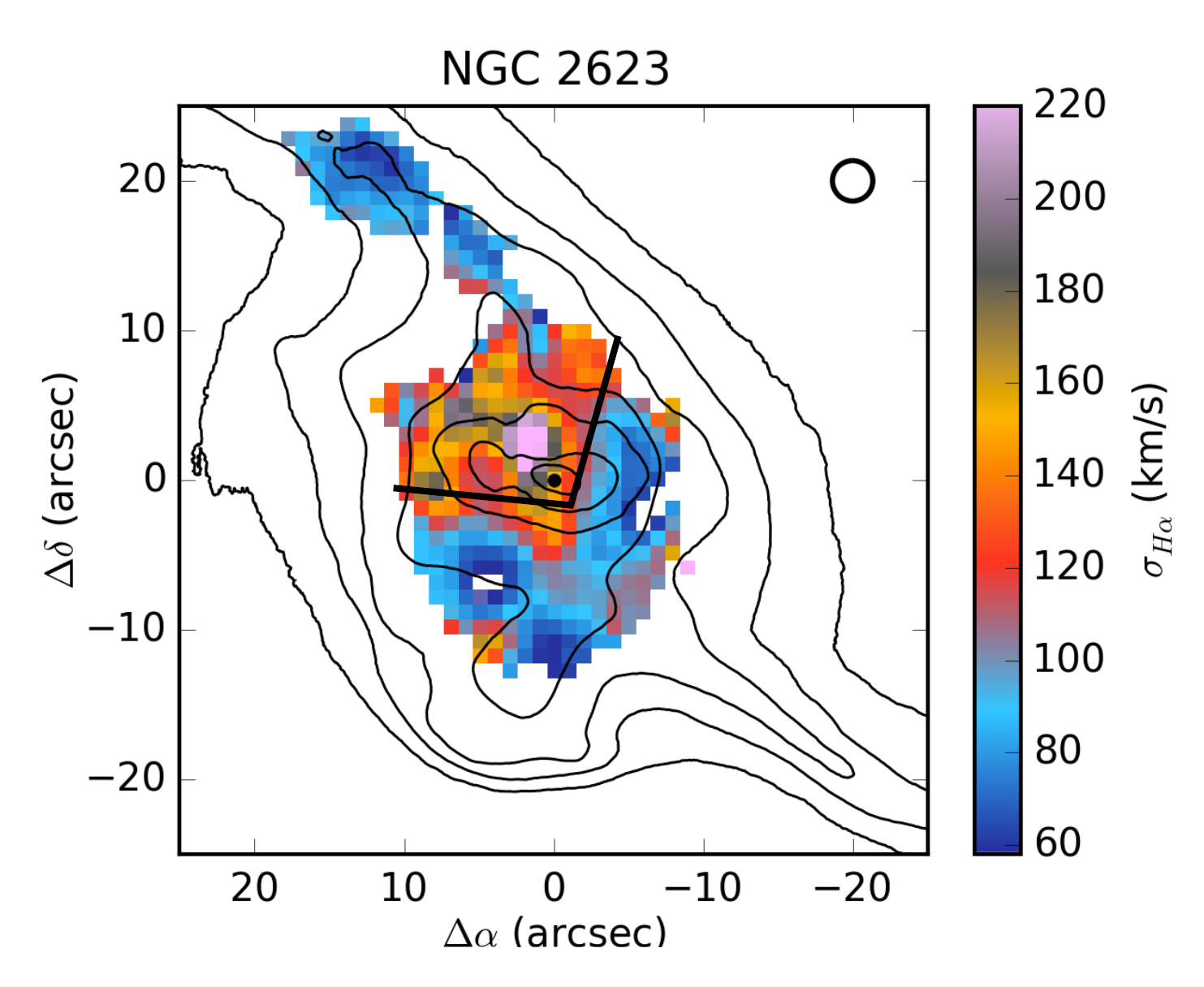}
\includegraphics[width=0.5\textwidth]{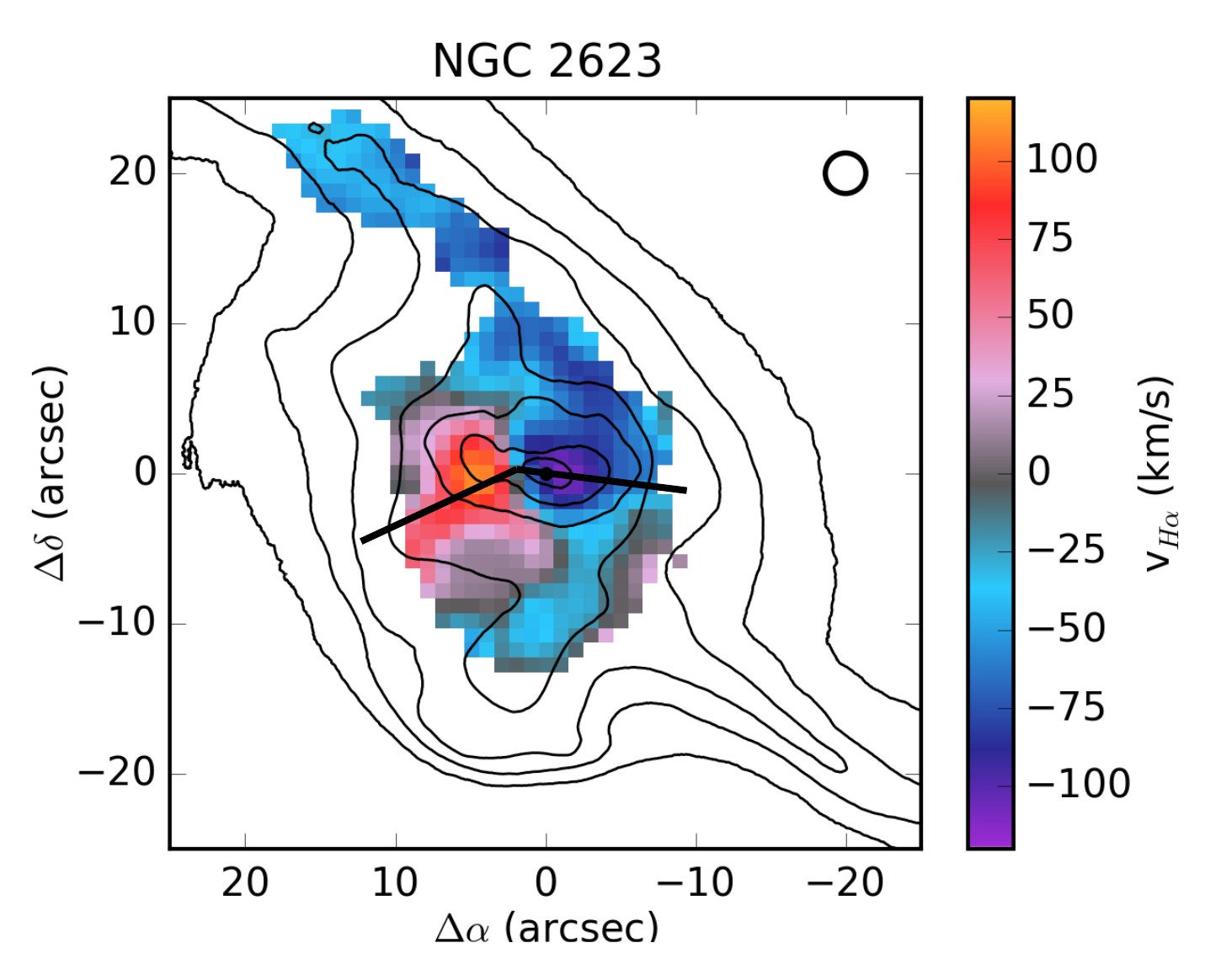}
\caption{Top panel: Velocity dispersion derived from H$\alpha$, 
corrected for the instrumental 
width ($\sigma_{inst}$ $\sim$ 116 km s$^{-1}$).
Bottom panel: Velocity field derived from H$\alpha$. 
The position angles of the kinematic major axis for the 
approaching a receding sides have been calculated by 
\protect\cite{barrera-ballesteros2015} and are shown in the 
figure with black solid lines. The circle in the top right 
corner of both maps represents the spatial resolution, given by 
the diameter of PPaK fibres, $\sim$ 2.7$\tt{''}$.}
\label{Fig_4p32}  
\end{center}
\end{figure}

\subsection{OSIRIS line emission at the edge of the SW tidal tail}\label{4.7}
The OSIRIS narrow-band filter imaging data 
covers a very large FoV, including 
the full extent of NGC 2623, beyond the tidal tails. 
In the OSIRIS 6720$\AA$ and 6725$\AA$ filter maps 
(6580 and 6585$\AA$ rest-frame at NGC 2623 centre),
we detect four regions 
located more than 19 kpc away from the centre, 
at the edge of the southwestern tidal tail.

As mentioned in Section \ref{2.2}, the wavelength tuning
is not uniform over the full field of view of OSIRIS. 
There is a progressive increasing shift to the blue of 
the central wavelength ($\lambda_{0}$) as the distance (r) from 
the optical centre increases.
At the positions of the four extended regions in the OSIRIS FoV, 
the filter is sensitive to
emission at wavelengths ranging from
6553 to 6562 $\rm \AA$, corresponding to rest-frame [NII]$\lambda$6548 and
H$\alpha$ emission; the measured fluxes in these regions are in the range 
2.5--7.7 $\times$10$^{-16}$ erg s$^{-1}$ cm$^{-2}$. 
The line emission indicates that these are star forming regions, 
presumably young.
\begin{figure*}
\begin{center}
\includegraphics[width=0.9\textwidth]{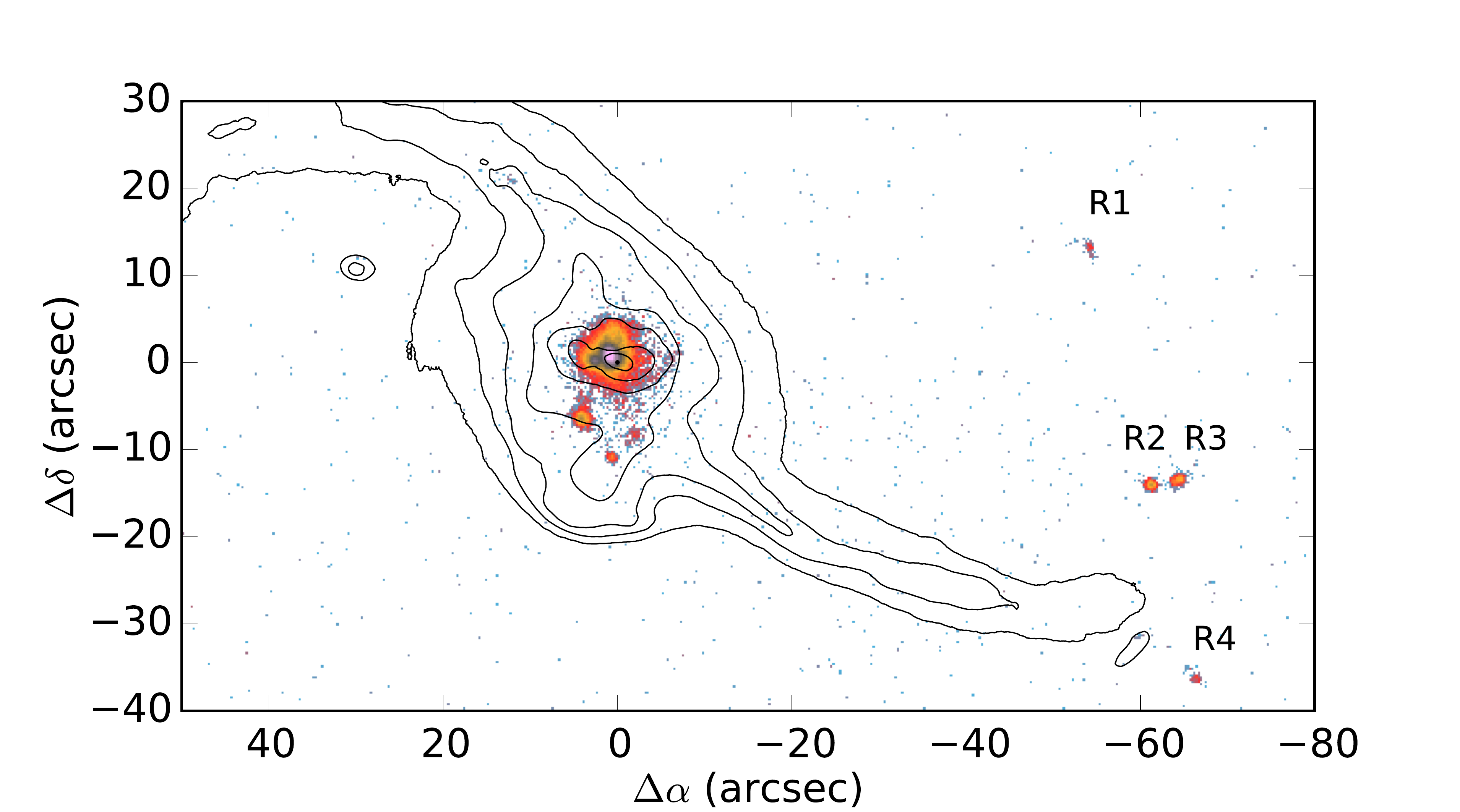}
\caption{OSIRIS $\sim$ 6723 $\AA$ map emission flux image derived from the OSIRIS data. This 
correspond to the [NII]$\lambda$6584 rest frame at the NGC 2623 centre, and to 
between [NII]$\lambda$6548 and H$\alpha$ emission in R1, R2, R3, and R4.}
\label{Fig_4p32}  
\end{center}
\end{figure*}

In Figure 16 we show the average 
of the OSIRIS 6720$\AA$ and 6725$\AA$ maps, 
where the star forming regions have been indicated as R1, R2, R3, and R4.
The latter were previously 
detected in H$\alpha$ and UV images of NGC 2623 by \citet{bournaud2004} and 
\citet{demello2012}, respectively.
We note that these regions do not coincide with the optical emission 
from the southern tidal tail. Instead, they are embedded in the underlying 
southern HI tail, which is located far away from the stellar 
component of NGC 2623, as found by \cite{hibbard&yun1996}. 
According to the HI kinematics, the southern HI tail is bending towards 
the North, incorporating the 4 star-forming regions. The two brigthest (R2 and R3) 
are near to HI emission peak, as can be seen in \cite{demello2012}. 
A similar case is given by the late-stage merger NGC 7252 \citep{lelli2015}, where 
there are two genuine tidal dwarf galaxies (TDGs) in formation. 
Through a detailed analysis of the gas dynamics and metallicty 
in this system, using HI and optical IFS observations, 
they reinforce that the TDGs in NGC 7252 were not pre-existing dwarfs, 
but currently forming out of pre-enriched material ejected from the 
progenitor galaxies. Morever, they found evidence of TDGs being 
devoided of dark matter.

Given the location of R1--R4 star forming regions in NGC 2623, 
they are compatible with the formation of
a tidal dwarf galaxy in the south-west tail of NGC 2623.
Unfortunately, we can not characterize them further, because these 
regions are not covered by the CALIFA 
data cube, so we cannot estimate their kinematics, 
and precise masses, ages, and metallicities.


\section{Discussion}\label{5}
Interactions, and in particular mergers, enhance the star formation, 
leading to star formation rates (SFRs) that are temporarily increased, 
sometimes by large amounts. Evolved mergers like NGC 2623, are ideal 
systems where to characterize, and quantify the merger-induced 
star formation, to understand the evolutionary path that major mergers 
follow. 
In this section we discuss our main findings in the context of 
the merger-driven evolution of galaxies.

\subsection{Global enhancement of the star formation}\label{5.1}

First, we calculate and compare the current star formation rate (SFR) 
of NGC 2623, using the spectral synthesis, Spitzer 24$\mu$m, 
FIR and radio continuum data.

The current SFR in different timescales ($t_{SF}$  = 30 Myr, 1 Gyr) is 
obtained using the SFH derived from the stellar continuum decomposition, 
by adding all the stellar mass formed in the last $t_{SF}$, 
and dividing it by this time.
We also used public IR data (MIPS 24$\mu$m from Spitzer, 
and IRAS 25, 60, and 100 $\mu$m) and radio 
continuum (Very Large Array, VLA at 1.4 GHz) to derive the current 
average SFR. Using \cite{kennicutt2009} hybrid calibrations, 
we found that the three of them lead to an average 
SFR $\sim$ 18 M$_{\odot}$ yr$^{-1}$, with a small dispersion 
of 1 M$_{\odot}$ yr$^{-1}$. This value is similar to SFR estimated from the 
spectral synthesis for $t_{SF}$  = 1 Gyr (SFR $\sim$ 12 M$_{\odot}$ yr$^{-1}$), 
and a factor 2 above if the SFR is calculated in the 
very recent past ($t_{SF}$  = 30 Myr) (SFR $\sim$ 8 M$_{\odot}$ yr$^{-1}$).
The difference in the SFR calculated with different indicators 
is a consequence of the SFR not being constant in the 
recent past of this galaxy, as the star formation history shows. 
However, the observed discrepancies are within the dispersion of
$\sim$0.2 dex found by \cite{pereira2015} when they compare the SFR derived 
from the FIR emission with SFR from the star formation history 
in time scale of $\sim$100 Myr in a sample of LIRGs.

These calculations show that the SFR in NGC 2623 is enhanced by a 
factor 2--3  with respect to main sequence star forming 
galaxies \citep{gonzalezdelgado2016}, and similar to the global 
enhancement of the star formation found in the pre-merger LIRGs 
IC 1623 and NGC 6090 (\citetalias{cortijo17}).

\subsection{Merger-induced star formation and evolutionary scheme}\label{5.2}

Another important aspect of mergers that must be taken into 
account is the spatial extension of the star formation. 
We have used multi-wavelength imaging from FUV to NIR 
to characterize the star cluster properties in NGC 2623 (see Appendix A).
Our analysis shows that NGC 2623 has experienced at least
two recent major episodes of massive cluster formation: 
one in the innermost nuclear regions 
(within 0.5 HLR $\sim$ 1.4 kpc) 
in the last $\lesssim$ 100 Myr, 
and other $\sim$ 250 Myr ago in the off-nuclear regions and 
some clusters in the circum-nuclear region (Section \ref{A.1}).
When considering only the optical colours, our results are 
in agreement with \cite{evans2008}.
However, we find that the age estimates can be improved by 
including the UV data.

Moreover, the cluster ages derived by us 
are consistent with the first dynamical simulation for 
NGC 2623 developed by \cite{privon2013}, predicting that 
first peri-centre passage occurred 
220 Myr ago, while the coalescence occurred 80 Myr ago.
Similar results are found in other evolved mergers, as 
Arp 220 and NGC 7252 \citep{miller1997,wilson2006}, 
while in contrast, in early-stage mergers as 
the Antennae, IC 1623 and NGC 6090, 
the young stellar populations are not so centrally concentrated, 
but widespread over several kpc scales \citep{zhang2001,cortijo17}.

From our IFS, we have also found that young stellar 
populations ($<$ 140 Myr) are mainly located 
in the innermost ($<$ 0.5 HLR $\sim$ 1.4 kpc) 
central regions; while a previous and 
widespread ($\sim$ 2 HLR $\sim$ 5.6 kpc) episode is traced 
by the spatially extended intermediate-age stellar 
populations with ages between 140 Myr--1.4 Gyr (Fig. 7). 
Due to this distribution of the stellar populations, the centre of 
NGC 2623 ($<$0.2 HLR) is younger, 500 Myr on average,
than at 1 HLR, $\sim$ 900 Myr (Fig. 6), and the age gradient 
in the inner 1 HLR, $\Delta_{in} age$, 
is positive ($\sim$ 400 Myr), in constrast with the negative 
gradients in Sbc--Sc galaxies ($-2.4$ Gyr/$-670$ Myr). 
The stellar extinction is high in the inner 0.2 HLR ($\sim$ 1.4 mag), 
and shows a negative gradient, which is much steeper (-0.9 mag)
than in Sbc-Sc galaxies, -0.3 mag (Fig. 5).

Other studies based on long-slit spectroscopy have found a 
similar distribution of the stellar populations in ULIRGs, 
including also the prototypical Arp 220 \citep{rodzau2008,rodzau2009}. 

The enhancement of the star formation, and the spatial distribution 
of the stellar populations in NGC 2623 support the evolutionary scheme 
proposed in \citetalias{cortijo17} for two early-stage mergers. In this scheme 
some star formation is initially triggered 
in the central regions during the first contact stage, 
followed by extended star formation after the first peri-centre passage.
In NGC 2623, we are observing the fossil record 
of this extended population (which is already a few 100 Myr to 
1 Gyr old), a relic of when NGC 2623 was at an earlier merger stage.
This result is also in agreement with high resolution 
models in the literature \citep{teyssier2010,hopkins2013,renaud2015}, 
which apart from the extended star-formation also predict a later 
nuclear starburst between the second peri-centre passage 
and the final coalescence, 
which is exactly what we find in NGC 2623.

We emphasise that the onset and evolution of star formation in mergers 
requires further observational studies. Not all early-stage mergers 
present extended and significant young star formation, as is the 
case of the post first passage Mice galaxies \citep{wild2014}. 
In a future paper we aim to extend the sample to cover more 
early-stage merger and merger systems, in order to 
observationally trace the star formation 
in mergers and put further constraints on the simulations.

\subsection{Ionized gas distribution and ionization mechanism}\label{5.3}

Most of the ionized gas in NGC 2623 is concentrated 
in the nuclear regions ($\lesssim$ 1 HLR), in contrast with the 
two early-stage merger LIRGs studied in \citetalias{cortijo17}, 
where the young stellar components are more widespread. 
This reinforces the idea that the young stellar 
components are mostly located in the central regions 
for advanced, post-coalescence systems.

This result found for NGC 2623 is supported by  
\cite{hattori2004}, who found that in separated pairs 
the contribution of extended starbursts ($>$ 1 kpc) to 
the H$\alpha$ emission is important, while 
in mergers (single coalesced nucleus) the
nuclear starbursts are very compact 
(from 100 pc--1 kpc) with no, or negligible, 
star forming activity in the outer regions. 
Analogously, \citet{garcia-marin2009,arribas2012} 
found that in the pre-coalescence systems the 
size of the H$\alpha$ emitting region is above 
2 kpc and up to 7--8 kpc, 
while for the post-coalescence systems it is $<$2 kpc.

From the diagnostic diagrams we find that only 
the line ratios of gas in the northern tidal tail and in 
the brightest knot south of the nucleus are consistent with
pure stellar photoionization. 
In the outflow cone nebula we find LINER-like ionization, 
together with the highest values of the 
velocity dispersion $\sim$ 220 km/s.
An ensemble of HII regions alone would not be able to reproduce 
the line ratios we observe, and shocks contribution from the outflow is 
necessary. However, we find that a collection of star forming 
regions is also present, and coexist with the outflow in the plane 
of the galaxy. Extended shock excitation has been previously 
found in many local U/LIRGs.
Closely interacting pairs and coalesced mergers show a high 
velocity dispersion component, above 100 km s$^{-1}$, and an increasingly 
dominant contribution from composite and 
LINER-like emission line ratios, tracing an increasing fractional 
contribution from shocks, exceeding half of the 
H$\alpha$ luminosity in late-stage coalesced mergers \citep{rich2015}.

\subsection{Intensity of the SFR and outflow escape}\label{5.4}

Using the H$\alpha$ luminosity of the shocked 
regions (i.e. spaxels with LINER-like ionization, 
L$^{shock}_{H\alpha}$ $\lesssim$ 2.5$\times$10$^{41}$ erg s$^{-1}$) 
we made an independent estimation of the current 
star formation rate which would be necessary 
to power the outflow. 
We have converted L$^{shock}_{H\alpha}$ into a bolometric 
energy loss rate  
$\dot{E}_{shock} \sim$ 2.0$\times$10$^{43}$ erg s$^{-1}$,
\footnote{Assuming that the bolometric conversion
factor for the shocks in NGC 2623 ($\sim$ 400 km/s) can be approximated by 
that computed by \cite{rich2010} for low velocity shock models, 
which tends to $\sim$80 for v$_{shock} =$ 140--200 km s$^{-1}$.} 
in order to compare with the mechanical energy released to the 
ISM by SNe from evolutionary synthesis models \citep{leitherer1999}. 
Assuming a continuous SFR, with a Salpeter IMF, and 
an average metallicity of $\sim$0.7 Z$_{\odot}$ (estimated 
from the spectral synthesis), we find that a SFR of 
$\lesssim$ 26 M$_{\odot}$ yr$^{-1}$ is 
required to power the observed outflow. This value is consistent 
with the SFR predicted from multi-wavelength 
data ($\sim$ 18 M$_{\odot}$ yr$^{-1}$), and represents 
an upper limit because some star formation contribution 
to the H$\alpha$ emission is also expected.

It is also interesting to know if NGC 2623 outflow 
is able to escape the galaxy and contribute to the ionization/enrichment 
of the intergalactic medium.
In principle, it seems unlikely given the low SFR intensity of 
$\sim$ 0.5 M$_{\odot}$ yr$^{-1}$ kpc$^{-2}$ of NGC 2623.
However, \cite{arribas2014} found that, while high-$z$ galaxies seem to require 
$\Sigma_{SFR}$ > 1 M$_{\odot}$ yr$^{-1}$ kpc$^{-2}$ 
to launch strong outflows, this threshold is not observed 
in low-$z$ U/LIRGs.
  
Following \cite{colina1991}, we estimate the total amount of ionized 
outflowing gas from the L$^{shock}_{H\alpha}$ above, and 
the electron density estimated in the cone-nebula $N_{e} \sim$ 100 cm$^{-3}$, 
which leads to $M_{g} \sim$ 1.1 $\times$10$^{7}$ M$_{\odot}$ (Chabrier IMF). 
\citetalias{lipari2004} derived an outflow velocity of 
$\langle V_{OF} \rangle$ $\sim$ $-405$ km s$^{-1}$, that, 
together with the radius 
of the outflowing region in kpc (r = 1.4 kpc), allows us to calculate 
the dynamical time, $t_{dyn} \sim$ 3.4 $\times$10$^{6}$ yr, and 
the outflow rate, $\dot{M_{g}} \sim$ 3.2 M$_{\odot}$/yr (Chabrier IMF), 
which is a factor 3 lower than the SFR ($\sim$ 11 M$_{\odot}$/yr, Chabrier IMF).
We also find that the outflow velocity is smaller than 
the escape velocity in the central areas, with an 
average $V_{esc} \sim$ 570 $\pm$ 70 km s$^{-1}$ 
\footnote{Using equation 7 of \cite{arribas2014}, 
assuming $r_{max} = 10$ kpc, and following equation 1 of 
\cite{bellocchi2013} to derive the dynamical mass of NGC 2623 inside 
the inner 1 half mass 
radius (2.2 kpc, $\sigma_{mean} \sim$ 131 km s$^{-1}$), which is
$\sim$ 5.3 $\times$10$^{10}$ M$_{\odot}$, with a factor 
1.3 of uncertainty, similar to the dynamical 
mass estimated by \cite{privon2013}.} $>$ $\langle V_{OF} \rangle$.  
These results indicate that in this case the outflow 
cannot escape and will be retained within the galaxy.

\subsection{NGC 2623 and the formation of an elliptical galaxy}\label{5.5}

Several studies proposed that major mergers will end up forming 
elliptical (E) galaxies. \cite{kormendy&sanders1992} were the first to propose 
that U/LIRGs will evolve into E through merger induced dissipative 
collapse. U/LIRGs have large molecular gas concentrations in their 
central kpc regions \citep{downes&solomon1998} with densities comparable to 
stellar densities in E.
They lie on the fundamental plane in the same region as intermediate 
mass E \citep{genzel2001}, and their near-IR brightness 
profiles follow a de Vaucouleurs law $r^{1/4}$ 
law \citep{wright1990,scoville2000}.

There is also evidence in the literature that NGC 2623 can evolve into an 
elliptical galaxy. They come from: a) the average rotation curve, which 
is well fitted by a Plummer spherical potential (\citetalias{lipari2004}); 
b) the NIR light surface brightness profile of the central 
region, which is well fitted by r$^{1/4}$ law by \cite{scoville2000}, 
using NICMOS data. 
However, this last work also indicate that an exponential law can also 
fit well the NIR light distribution if the central 
region ($<$ 2 kpc)  is excluded.

Our analysis provides additional information about the 
future evolution of this LIRG/merger system. In Section \ref{3.2} we 
found that the radial distribution of the stellar mass surface density 
is quite similar to Sbc galaxies in CALIFA, which are significantly 
less dense than massive E galaxies, and its central 
gradient ($\Delta_{in}$ log $\mu_{\star}$ = $-1.2$) flatter 
than the gradient in massive (M$_\star > 10^{11}$ M$_\odot$) E--S0 galaxies 
($\Delta_{in}$ log $\mu_{\star}$ = $-1.4$, \citetalias{gonzalezdelgado2015}).
Because this difference in the gradient with respect to 
E--S0 galaxies could be due to an insufficient extinction correction or to the 
limited spatial resolution in the inner regions, we have proceeded as follow.
Using the SFH history derived from the spectral fits, we have estimated the 
synthetic radial profile 
of the 1.6$\mu$m luminosity surface density, and we have compared it with 
the NICMOS F160W luminosity surface 
brightness. We find that both radial profiles are very 
similar, and there exists only a discrepancy in the innermost 0.25 HLR, 
where the NIC2/NIC3 F160W profile is slightly steeper. This discrepancy is easily 
explained by the worse spatial resolution of our IFS data in 
comparison with NICMOS. This prevents us to trace with the IFS 
the inner steep rise of the radial profile, 
while outside of $>$0.7 HLR 
we find an exponential profile, in agreement 
with \cite{scoville2000} (see Appendix C and Fig. C.1).

As an additional check, we have explicitily extracted the 
mass surface density profiles of E/S0, Sa, and Sb galaxies 
in the CALIFA sample with stellar masses in the same mass range as 
NGC 2623 (3$\times 10^{10}$--8$\times 10^{10}$ $M_\odot$). 
The E/S0 sub-sample of intermediate mass are now more 
similar to NGC 2623, with a global difference 
in the mass surface density of $\sim$0.3 dex, and 
$\nabla_{in}$ log $\mu_{\star}$ = $-1.3$, 
which is closer to 
the $\nabla_{in}$ log $\mu_{\star}$ in NGC 2623. 
Thus, our conclusion is that if NGC 2623 forms 
an E, it has to be less 
massive than 10$^{11}$ M$_{\odot}$.
Alternatively, 
we cannot discard the possibility that NGC 2623 will 
end up as a spiral galaxy with a prominent 
bulge, because  $\nabla_{in}$ log $\mu_{\star}$ is 
also similar to Sa--Sb galaxies (Figure 17).
Recent simulations have found that gas-rich major mergers can also produce
spiral remnants. The stellar progenitor disks are first 
destroyed by violent relaxation during the mergers, 
forming classical bulges, 
but new disks soon regrow out of the leftover debris, as well as
gas accreted mainly from the haloes \citep{springel2005,athanassoula2016}.

\begin{figure}
\begin{center}
\includegraphics[width=0.5\textwidth]{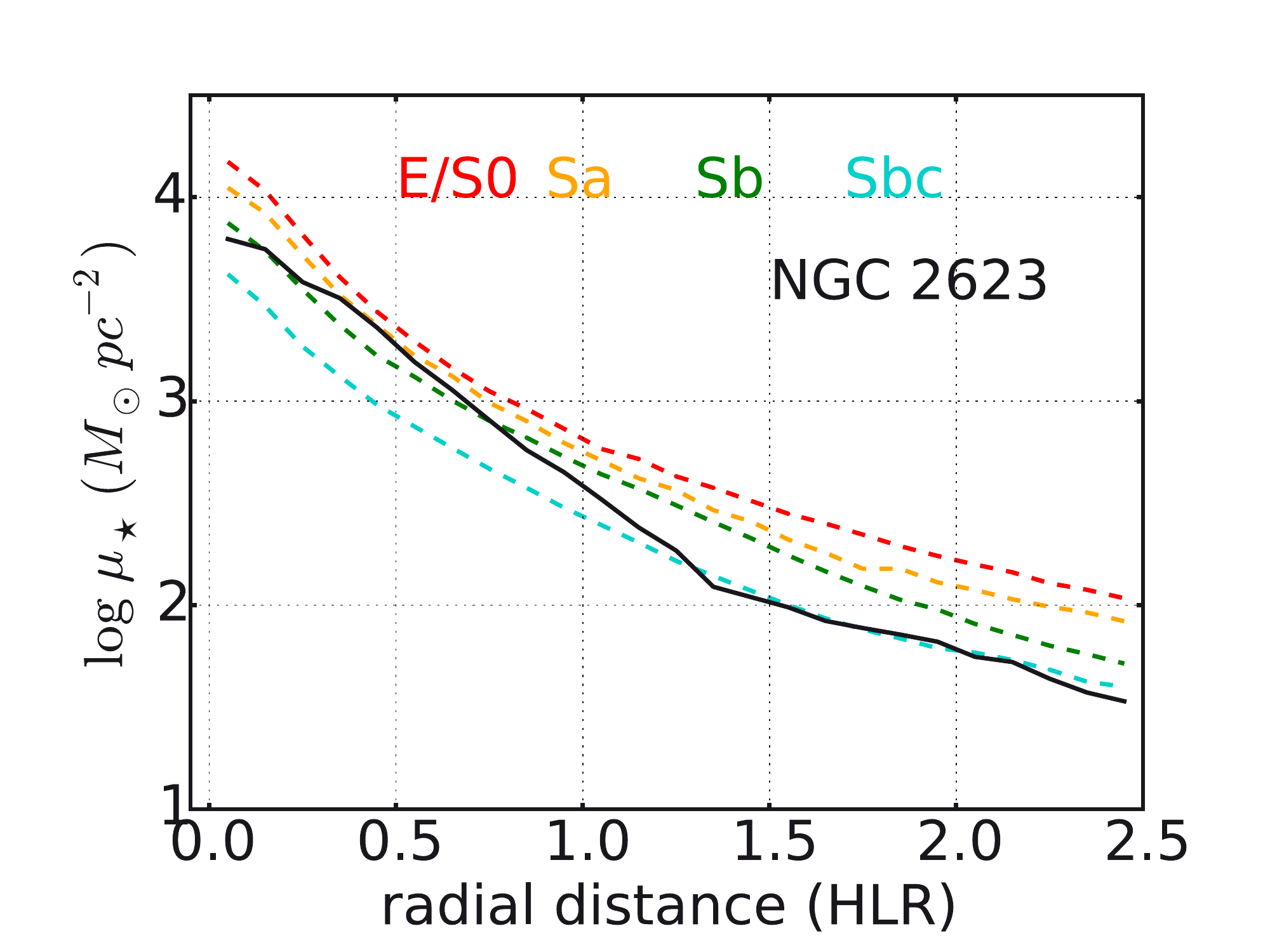}
\caption{Radial profile of $\log \ \mu_\star$ for 
NGC 2623 (black line) and for a sub-sample of E/S0 and spiral 
galaxies from the CALIFA survey that have a stellar 
mass ($10.5 \leq \log \ M_\star \ (M_\odot) \leq 10.9$) 
similar to NGC 2623. }
\label{Fig5p5_mu}  
\end{center}
\end{figure}

\section{Summary and conclusions}
\label{6}
We have characterized the star cluster properties, 
the stellar population and ionized gas properties, 
and the ionization mechanism in the merger LIRG NGC 2623, 
by analysing high quality IFS data, 
from PMAS LArr and CALIFA,
in the rest-frame optical range 3700--7000 \AA, 
high resolution HST imaging, and OSIRIS narrow band H$\alpha$ and 
[NII]$\lambda$6584 imaging. 
NGC 2623 results have been compared 
with control Sbc and Sc galaxies from CALIFA (\citetalias{gonzalezdelgado2015}) in the same mass range, 
and with the early-stage merger LIRGs IC 1623 W 
and NGC 6090 studied in \citetalias{cortijo17}. 
Our main results are:

\begin{itemize}
\item[-] From the IFS and clusters photometry we find 
two periods of merger-induced star formation in NGC 2623: a recent episode, 
traced by young stellar populations, YSP ($<$ 140 Myr), and located 
in the innermost ($<$ 0.5 HLR $\sim$ 1.4 kpc) 
central regions, and in some isolated clusters 
in the northern tidal tail and to the south 
of the nucleus; and an earlier and  more
widespread ($\sim$ 2 HLR $\sim$ 5.6 kpc) episode, traced 
by the spatially extended intermediate-age stellar 
populations, ISP (between 140 Myr--1.4 Gyr). 
This result is consistent with 
the epochs of the first peri-center passage ($\sim$200 Myr ago) 
and coalescence ($<$100 Myr ago) predicted by dynamical models, 
and in agreement with high resolution merger simulations in the literature.

\item[-] The central region (inner 0.2 HLR) shows the highest 
extinction with $A_{V}^{stars}$ = 1.4 mag. 
Moreover, in the inner 1 HLR, there exists a negative $A_{V}^{stars}$ gradient 
which is much steeper in NGC 2623 ($-0.9$ mag) than in Sbc/Sc galaxies ($-0.3$ mag). 
This is in contrast to the flat $A_{V}^{stars}$ profiles in 
the early-stage merger LIRGs, 
and is consistent with the idea that, in mergers with coalesced nuclei,
most of the dust content is already concentrated in the central regions.

\item[-] The age gradient of NGC 2623 in the inner 1 HLR, $\Delta_{in} age$, 
is positive ($\sim$ 400 Myr), in contrast to the negative 
gradient found in Sbc/Sc galaxies ($-2.4$ Gyr/$-670$ Myr).
The positive age gradient could be explained in advanced mergers 
if most of the gas/star formation is concentrated in the centre, while 
the outer parts evolve passively.

\item[-] Only gas in the northern tidal tail and the brightest knot south 
of the nucleus are purely ionized 
by young massive stars. In the nuclear region of NGC 2623, we find 
LINER-like ionization and high values 
of the velocity dispersion ($\sim$ 220 km s$^{-1}$), consistent with 
shock excitation associated with a known outflow.
As revealed by the highest resolution 
OSIRIS and HST data, a collection of HII regions is also present in the 
plane of the galaxy, which explain the mixture of ionization 
mechanisms in this system.

\item[-] The three hybrid tracers (observed H$\alpha$ from the IFS, 
together with 24$\mu$m, FIR, and radio 1.4 GHz), lead to 
an average SFR $\sim$ 18 M$_{\odot}$ yr$^{-1}$ (Salpeter IMF), 
similar to the one derived from the spectral synthesis, averaging 
in the last 1 Gyr, SFR($<$ 1 Gyr) $\sim$ 12 M$_{\odot}$ yr$^{-1}$.

\item[-] The low SFR intensity 
($\sim$ 0.5 M$_{\odot}$ yr$^{-1}$ kpc$^{-2}$), 
the fact that the outflow rate is 3 times 
lower than the current SFR, and the 
escape velocity in the central areas higher 
than the outflow velocity, all suggest that NGC 2623 outflow 
cannot escape and will be retained within the galaxy.

\item[-] The stellar mass density profile of 
NGC 2623 resembles that of Sbc galaxies in CALIFA sample, 
and it is 0.6 dex less dense, globally, than E--S0.
These results indicate that NGC 2623 may form a 
low-intermediate mass E (log $M_{\star} \lesssim$ 11) but 
not a very massive one, or, consistently with 
recent simulations, a spiral galaxy with a prominent 
bulge (Sa--Sbc).

\end{itemize}

\begin{acknowledgements} 
CALIFA is the first legacy survey carried out at Calar Alto. The CALIFA 
collaboration would like to thank the IAA-CSIC and MPIA-MPG as major partners 
of the observatory, and CAHA itself, for the unique access to telescope 
time and support in manpower and infrastructures. We also thank 
the CAHA staff for the dedication to this project. Support from the Spanish 
Ministerio de Econom\'ia y Competitividad, through projects AYA2016-77846-P, 
AYA2014-57490-P, AYA2010-15081, and Junta de Andaluc\'ia FQ1580.
ALdA, EADL, and RCF thank the hospitality of the 
IAA and the support of CAPES and CNPq. RGD acknowledges the support of 
CNPq (Brazil) through Programa Ci\^encia sem Fronteiras (401452/2012-3). 
MVM acknowledges support from Spanish MINECO through 
grant AYA2015-64346-C2-2-P. CCF acknowledges the constructive comments 
from Carlos L\'opez-Cob\'a and Salvador Duarte Puertas. 
This research made use of Python (http://www.python.org); 
Numpy \citep{vanderwalt2011}, and Matplotlib \citep{hunter2007}.
We thank the anonymous reviewer for his/her 
careful reading of our manuscript and the insightful comments 
and suggestions.
\end{acknowledgements}






\appendix
\section{Star cluster photometry}\label{A}
\begin{table*}
\centering
 \caption{Summary of HST data for NGC 2623}
 \label{tab:natbib}
 \begin{tabular}{ccccccccc}
  \hline
\hline

	&	    & Detector/  & Plate Scale          & Observation &t$_{exp}$  & Proposal ID,  \\ 
Filter & Instrument & Camera     & (arcsec pixel$^{-1}$)   & Date        & (s)     & PI  \\ 
 \hline
F140LP    & ACS	     & SBC      &  0.025      	    &2008-10-30  &2520	  & 11196, A. evans  \\
F330W     & ACS	     & HRC      &  0.025      	    &2004-12-28  &5465	  & 9735, B. Whitmore \\
F435W     & ACS	     & WFC      &  0.05       	    &2004-02-06  &3729.9  & 9735,B. Whitmore   \\
F435W     & ACS	     & WFC      &  0.05       	    &2005-11-29  &1275    & 10592, A. evans   \\
F555W     & ACS	     & WFC      &  0.05       	    &2004-02-06  &1206    & 9735,B. Whitmore  \\
F814W     & ACS	     & WFC      &  0.05       	    &2004-02-06  &2460    & 9735,B. Whitmore   \\
F814W     & ACS	     & WFC      &  0.05       	    &2005-11-29  &730     & 10592, A. evans   \\
F110W     & NICMOS   & NIC3     &  0.2        	    &2004-01-13  &2559.76 & 9735, B. Whitmore  \\
F160W     & NICMOS   & NIC3     &  0.2        	    &2004-01-13  &2559.74 & 9735, B. Whitmore  \\
\hline
 \end{tabular}
\end{table*}

\begin{figure*}  
\begin{center}
\includegraphics[width=\textwidth]{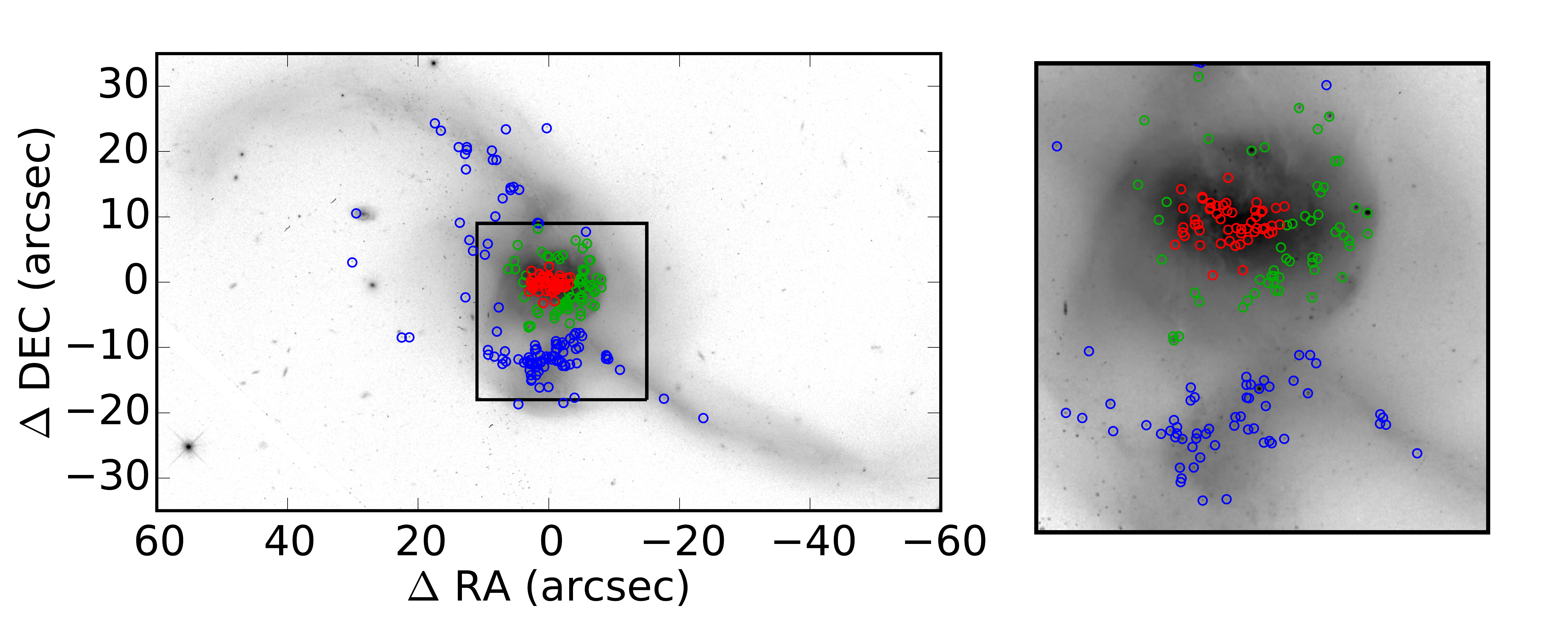}
\caption{F814W image of NGC 2623. North is up, East to left.  
The detected clusters are marked with apertures of different colours 
given their distances with respect to the centre: 
1) blue clusters are those with d $>$ 1 HLR, 
2) green clusters if $0.5$ HLR $< d \leq$ 1 HLR, and 
3) red clusters if d $<$ 0.5 HLR.
In the left panel the whole  galaxy is shown, where the black rectangle 
indicates the region simultaneously covered by all the 
filters from UV to NIR. 
A zoom into this region is shown in the right panel.}
\label{Fig_2}  
\end{center}
\end{figure*}

We have roughly estimated the stellar population 
properties of the super star clusters 
(SSCs) in NGC 2623 using the 
HST images from FUV to NIR. 
According to \cite{alvensleben2004}, a long wavelength 
basis from U band through NIR is necessary to 
determine the SPs properties of star clusters in a 
reliable way, where the availability 
of U band photometry is crucial. 
The characteristics of the HST images used by us are summarized 
in Table A.1. 
The  methodology followed for cluster detection and aperture 
photometry was described in Appendix A 
of \cite{cortijo17}.

By comparing the colours and magnitudes 
of NGC 2623 clusters with solar metallicity 
Charlot \& Bruzual(2007, unpublished) stellar population
synthesis models, we have derived their main properties; 
ages, masses, and the amount of dust obscuration. 
Model colours have been computed using STSDAS.SYNPHOT software. 

Clusters were detected in a combination of optical HST images with 
IRAF DAOFIND task, and photometry performed using the phot 
task in APPHOT. 

From 0--1 HLR there exists a strong negative 
$A_{V}$ gradient in NGC 2623 (see Figure 5). To take into account the 
extinction effects with distance, we have decided to separate the clusters 
into three bins, according to their distances (d) to the centre: 
\begin{enumerate}
\item \textbf{Blue clusters}: d $>$ 1 HLR
\item \textbf{Green clusters}: $0.5$ HLR $< d \leq$ 1 HLR
\item \textbf{Red clusters}: d $<$ 0.5 HLR
\end{enumerate} 
Basically, the blue clusters are located in the 
concentration of clusters south of the nucleus reported 
by \cite{evans2008}, while the red 
ones are located in the highly obscured nuclear 
region. Strong dust lanes can be seen in the nucleus
of NGC 2623 in HST B and V band images.

A total of 156 clusters where detected with SNR $>$ 5 
in FUV, F435W, and F814W, 
in the region in common for all filters.
Similarly, 203 clusters where detected with 
SNR $>$ 5 in F435W, F814W, and F160W.
The clusters are shown in Figure A1. 
In the left panel the whole galaxy is shown, where the black 
rectangle indicates the region simultaneously covered 
by all the filters from UV to NIR. 
A zoom into this region is shown in the right panel. 

\subsection{Clusters age and extinction}\label{A.1}
We have compared the colours and magnitudes of 
NGC 2623 clusters with 
those of Charlot \& Bruzual SSP models (2007, unpublished) 
in an age range 1 Myr to 13 Gyr.
Despite the recent
improvements in modeling techniques, it is very
difficult to estimate the metallicity of individual clusters from photometry.
From the spectroscopy we obtain an average value for the metallicity 
of $\sim$ 0.7 Z$_{\odot}$. 
Taking this into account, we have used the Z$_{\odot}$ SSP models as 
a proxy in this analysis.

Comparing the position of the clusters in colour-colour diagrams
with respect to the model positions we can estimate 
the cluster ages. In Figure A2 we show the FUV - F435W vs 
F435W - F814W diagram, which is the best to break 
the age - A$_V$ degeneracy.
The colour coding is the same as in Figure A1,
but only those clusters with SNR $\geq$ 5 
in FUV, F435W, and F814W filters are included in the plots.
The black solid line is the path described by SSP models
from 1 Myr to 13 Gyr, Z$_{\odot}$, and A$_{V}$ = 0 mag. 
The greyscale lines are the paths 
for the same models reddened by 1 to 3 mag, 
with the lighter shades tracing the more reddened models.

Similarly, in Figure A3 we show the F435W - F814W vs 
F814W - F160W diagram. Only clusters with SNR $\geq$ 5 
in F435W, F814W, and F160W filters are shown.
Again, the black solid line is the path described 
by SSPs from 1 Myr to 13 Gyr, Z$_{\odot}$, and 
A$_{V}$ = 0 mag. The greyscale lines are the paths 
for the same models reddened by up to 6 mag, with the 
lighter shades tracing the larger A$_{V}$ models.

From one side, the concentration of clusters south of the 
nucleus (d $>$ 1 HLR)
is consistent with not being affected by foreground extinction. 
They have ages ranging 100--400 Myr, with an 
average around $\sim$ 250 Myr.
Only a few of them could be younger if we consider they
are affected by some extinction.
On the contrary, the nuclear clusters 
(d $<$ 0.5 HLR) 
are affected by high 
extinction,  1--3 mag. 
Most of them consistent with ages younger than 100 Myr. 
This suggests that the nuclear region of NGC 2623 is 
slightly younger than its surroundings. 
The greater scatter is in the circumnuclear clusters 
$0.5$ HLR $< d \leq$ 1 HLR. 
They are consistent with an 
average extinction of around 1 mag. Given their spread in 
the diagrams they are compatible with having ages 
ranging 1--500 Myr.
We note that the photometric estimation of the 
stellar population properties, although rougher, is consistent 
with the spectroscopic one.
\begin{figure}
\begin{center}
\includegraphics[width=0.5\textwidth]{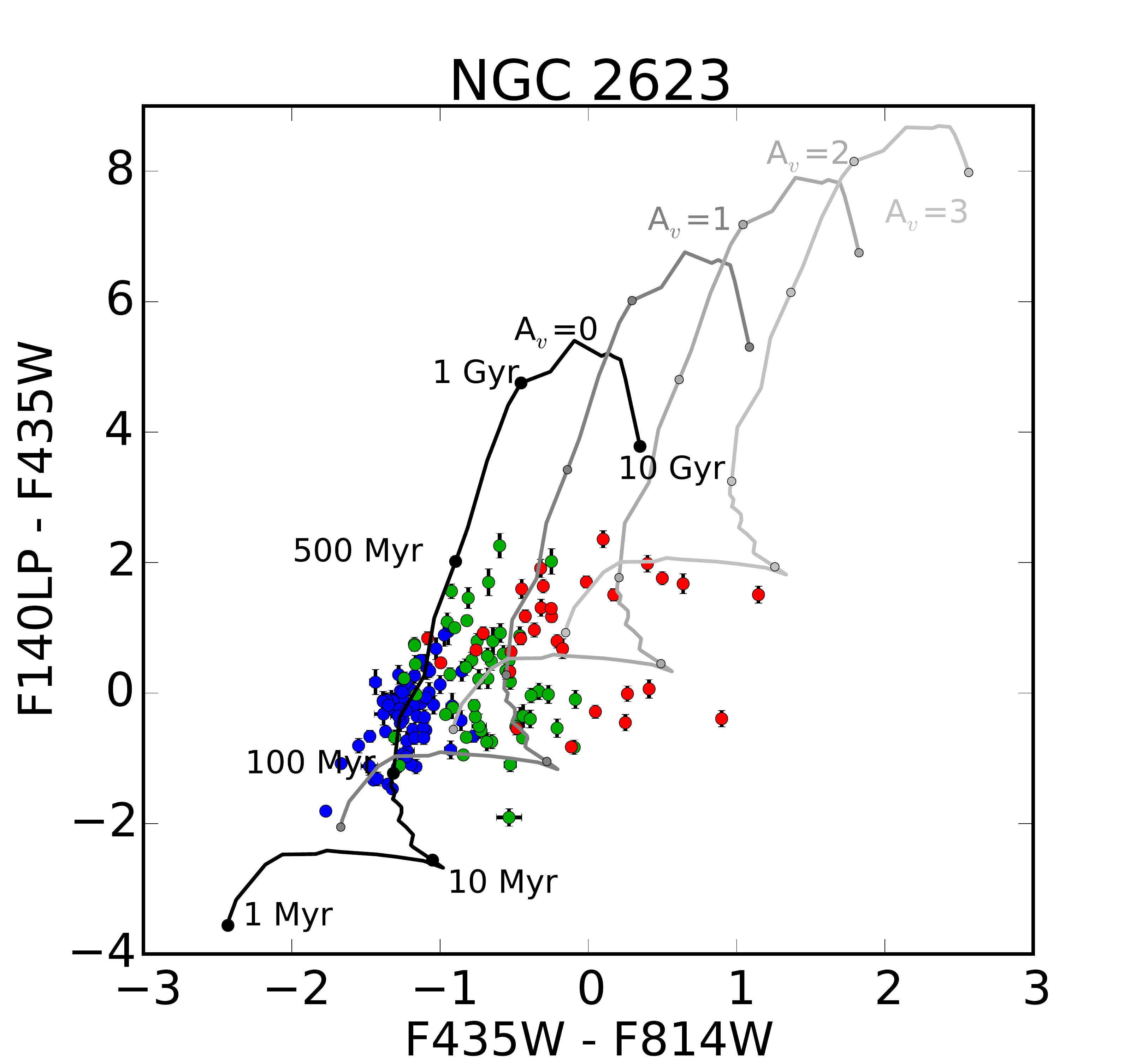} 
\caption{FUV - F435W vs F435W - F814W diagram for NGC 2623 clusters.
The color coding is the same as in Figure A1, 
but only those clusters with SNR $>$ 5 in FUV, F435W, and F814W filters 
are included in the plots. The black solid line is the path described by SSPs 
from 1 Myr to 13 Gyr, Z$_{\odot}$, and A$_{V}$=0 mag. The greyscale 
lines are the paths for the same models reddened by 1 to 3 mag, 
with the lighter shades tracing the more extincted models.}
\label{Fig_5p3}  
\end{center}
\end{figure} 

\subsection{Clusters masses}\label{A.2}
We found that NICMOS 1.6 $\mu m$ band is the most sentitive 
to mass variations. 
We will estimate the mass range of the star clusters, by 
comparing their 1.6 $\mu m$ absolute magnitude with the 
absolute magnitude of SSP models of masses ranging from 10$^{4}$ 
to 10$^{8}$ M$_{\odot}$.
The results are shown in Figure A4.
This is an optical color-color vs NIR 1.6$\mu m$ absolute magnitude diagram.
The paths from models of 10$^{4}$, 10$^{5}$, 10$^{6}$, 10$^{7}$, and 
10$^{8}$ M$_{\odot}$ are shown as solid and dashed lines.
The models have been cut between 1--500 Myr, as we know from the
previous section that these are the older ages our clusters can have.
We found a large range of masses in the clusters of NGC 2623.
The intermediate-age clusters in the region south of the
nucleus (d $>$ 1 HLR, blue) have masses 
from 10$^{4}$--10$^{5}$ M$_{\odot}$.
For the circumnuclear ($0.5$ HLR $< d \leq$ 1 HLR, green)
the spread is larger 
10$^{4}$--10$^{7}$ M$_{\odot}$.
On the other hand, the vast majority of the inner nuclear 
clusters (d $<$ 0.5 HLR, red) are consistent 
with masses between 
10$^{6}$--10$^{8}$ M$_{\odot}$.
These impressively large masses are comparable to 
the most massive globular clusters 
seen in giant elliptical galaxies and the massive, 
intermediate-age clusters seen in NGC 7252 and 
NGC 1316 \citep{maraston2004,bastian2006b}, and
some YMC in Arp 220 \citep{wilson2006}. 
We think, however, that we are not measuring masses of individual 
clusters, but aggregates of them.
\begin{figure} 
\begin{center}
\includegraphics[width=0.5\textwidth]{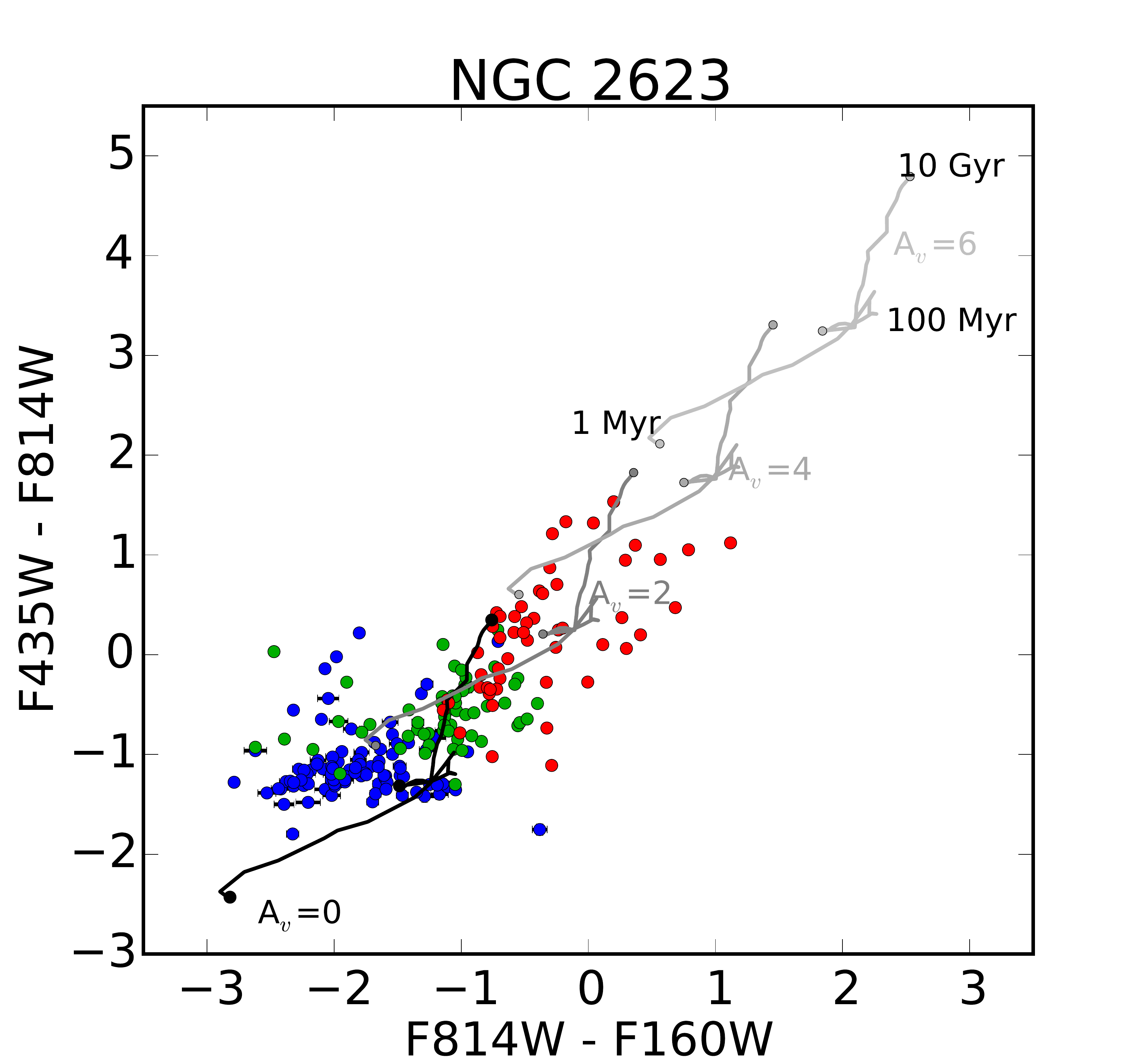} 
\caption{F435W - F814W vs F814W - F160W diagram for NGC 2623 clusters. 
Only clusters with SNR $>$ 5 in F435W, F814W, and F160W filters are shown.
The black solid line is the path described by SSPs 
from 1 Myr to 13 Gyr, Z$_{\odot}$, and A$_{V}$=0 mag. The greyscale lines 
are the paths for the same models but reddened by 2, 4, and 6 mag. The clusters 
color coding is the same as in Figure A1.}
\label{Fig_5p4}  
\end{center}
\end{figure} 
\begin{figure} 
\begin{center}
\includegraphics[width=0.5\textwidth]{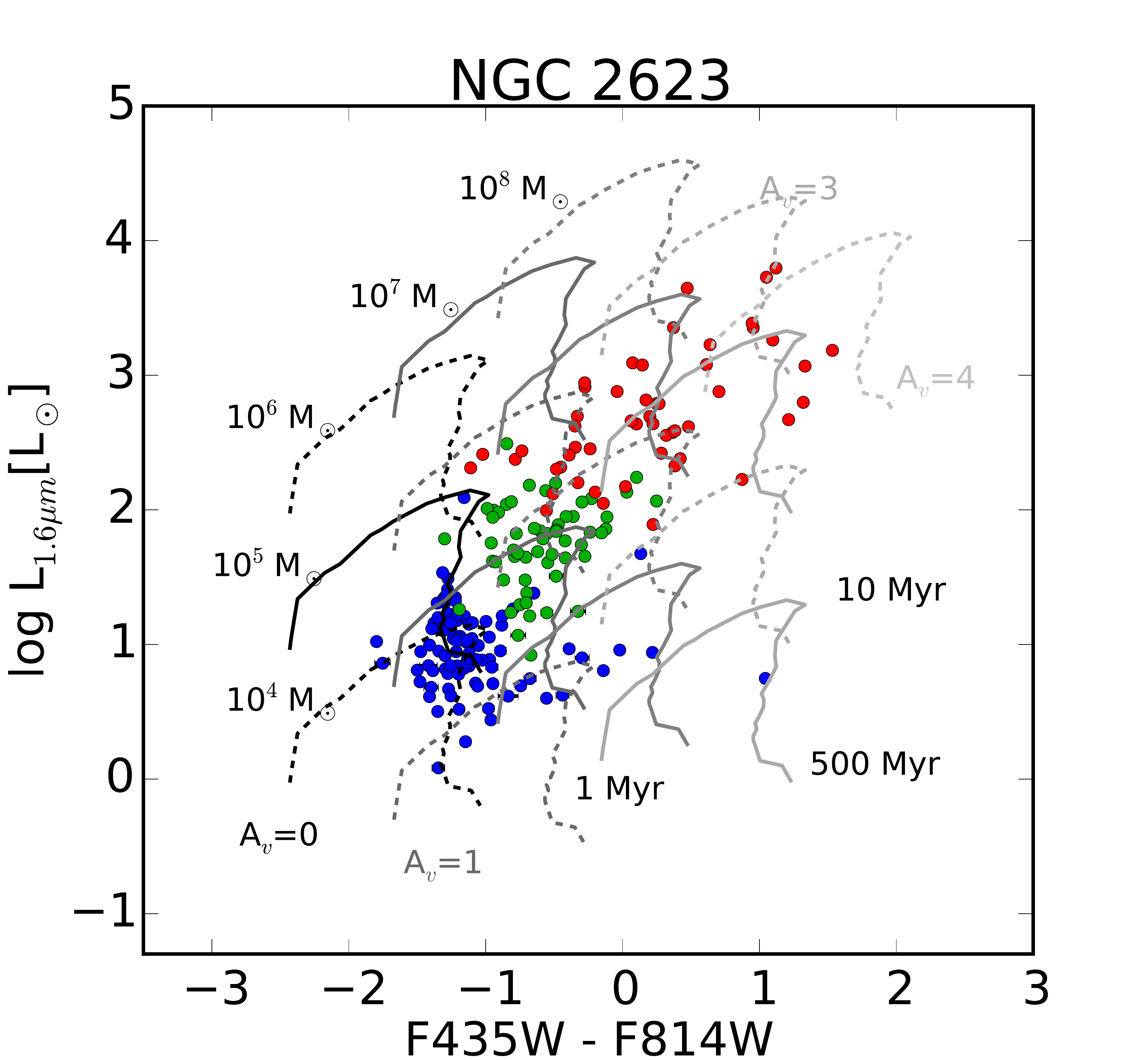} 
\caption{Right: NIC 1.6 $\mu$m absolute magnitude vs F435W - F814W diagram. 
Only clusters with SNR $>$ 5 in F435W, F814W and F160W filters are shown. 
The clusters color coding is the same as in Figure A1.
The solid and dashed lines are the path of SSPs from 1 Myr to 500 Myr and 
with masses from 10$^{5}$ to 10$^{8}$ M$_{\odot}$. The greyscale coding 
represents the variation in stellar extinction as in previous figures.}
\label{Fig_5p5}  
\end{center}
\end{figure}

We have performed a rough estimate of the total mass 
in clusters in NGC 2623, by assigning a certain 
average mass for the clusters in each luminosity 
L$_{1.6 \mu m}$ range. In particular, we have 
separated the clusters in five possible 
luminosity ranges: if log L$_{1.6 \mu m}$ $<$ 0.9 
we assign a mass of $\sim$ 10$^{4}$ M$_{\odot}$,
from 0.9 $<$ log L$_{1.6 \mu m}$ $<$ 1.5 
we assign a mass of $\sim$ 10$^{5}$ M$_{\odot}$, 
if 1.5 $<$ log L$_{1.6 \mu m}$ $<$ 2.65 
then $\sim$ 10$^{6}$ M$_{\odot}$, 
if 2.65 $<$ log L$_{1.6 \mu m}$ $<$ 3.5 
then $\sim$ 10$^{7}$ M$_{\odot}$, and 
above log L$_{1.6 \mu m}$ $>$ 3.5 we assign
$\sim$ 10$^{8}$ M$_{\odot}$. 
In the case of NGC 2623 we have 
39, 64, 75, 22, and 3 clusters in each range.
That leads to a rough estimate of the total mass 
in clusters in NGC 2623 of 
M$_{NGC 2623} ^{clus}$ $\sim$ 6 $\times$ 10$^{8}$ M$_{\odot}$, 
which represents 1$\%$ of the total stellar mass.

The mass in young components derived through the IFS is comparable 
(only a factor $\sim$ 1.5 less) to the mass in star clusters derived 
roughly from the photometry. 
This would agree with the hypothesis that the vast majority 
of stars form in clusters rather than in isolation. 

\subsection{Cluster detection limits}\label{A.3}
As explained in \citetalias{cortijo17}, 
a major question to be addressed with star cluster photometry 
is to know whether older/more reddened clusters could have been detected.
We followed a simplistic approach to estimate 
which clusters could be, in principle, detected, and which not.
Using the SSP models of 10 Myr, 100 Myr, 200 Myr, 300 Myr, 500 Myr, 
and 1 Gyr, and the transmission curves from F435W and F814W filters, 
we compute the 435 and 814 magnitudes for clusters of different masses 
($10^5 M_{\odot}$, $10^6 M_{\odot}$, $10^7 M_{\odot}$, 
and $10^8 M_{\odot}$) and affected by extinctions of 
0 mag, 1 mag, 2 mag, 4 mag, 6 mag, and 10 mag. 
These magnitudes are compared with the F435W and F814W limiting magnitudes 
expected for point sources (most clusters are not resolved) with 
SNR=5, obtained from Figure 10.2 and 10.32 of the ACS Instrument Handbook, 
given the exposure times of the HST observations used in 
this paper ($\sim$ $10^{3}$ s). 

Given the fact that NGC 2623 is nearer, 
but with a distance very similar to IC 1623 W (80 Mpc vs 86.8 Mpc), 
the detection limits for IC 1623 W apply to NGC 2623 as well. 
The less massive the cluster the more difficult to 
detect it as it ages or is more reddened. 
Clusters of $10^5 M_{\odot}$ up to 300 Myr can be detected 
for no extinction, for $A_{V}$ = 1 mag, clusters older than 100 Myr 
cannot be detected, and for $A_{V}$ = 2--4 mag, only clusters of 
10 Myr can be detected. 
$10^6 M_{\odot}$ clusters up to 1 Gyr could be detected up to 
extinction levels of $A_{V} = 2$ mag. 
For $A_{V} = 4$ mag, only clusters up to 300 Myr can be detected, 
and for $A_{V} = 6$ mag, only 10 Myr clusters can be detected. 
Above that extinction level, 
we cannot detect clusters anymore. 
More massive clusters, $10^{7}-10^{8} M_{\odot}$, can be 
detected without problems up to ages of 1 Gyr, 
and $A_{V} = 6$ mag.

\section{Uncertainties related with models choice: GM vs. CB bases}\label{B}
\begin{figure}
\begin{center}
\includegraphics[width=0.47\textwidth]{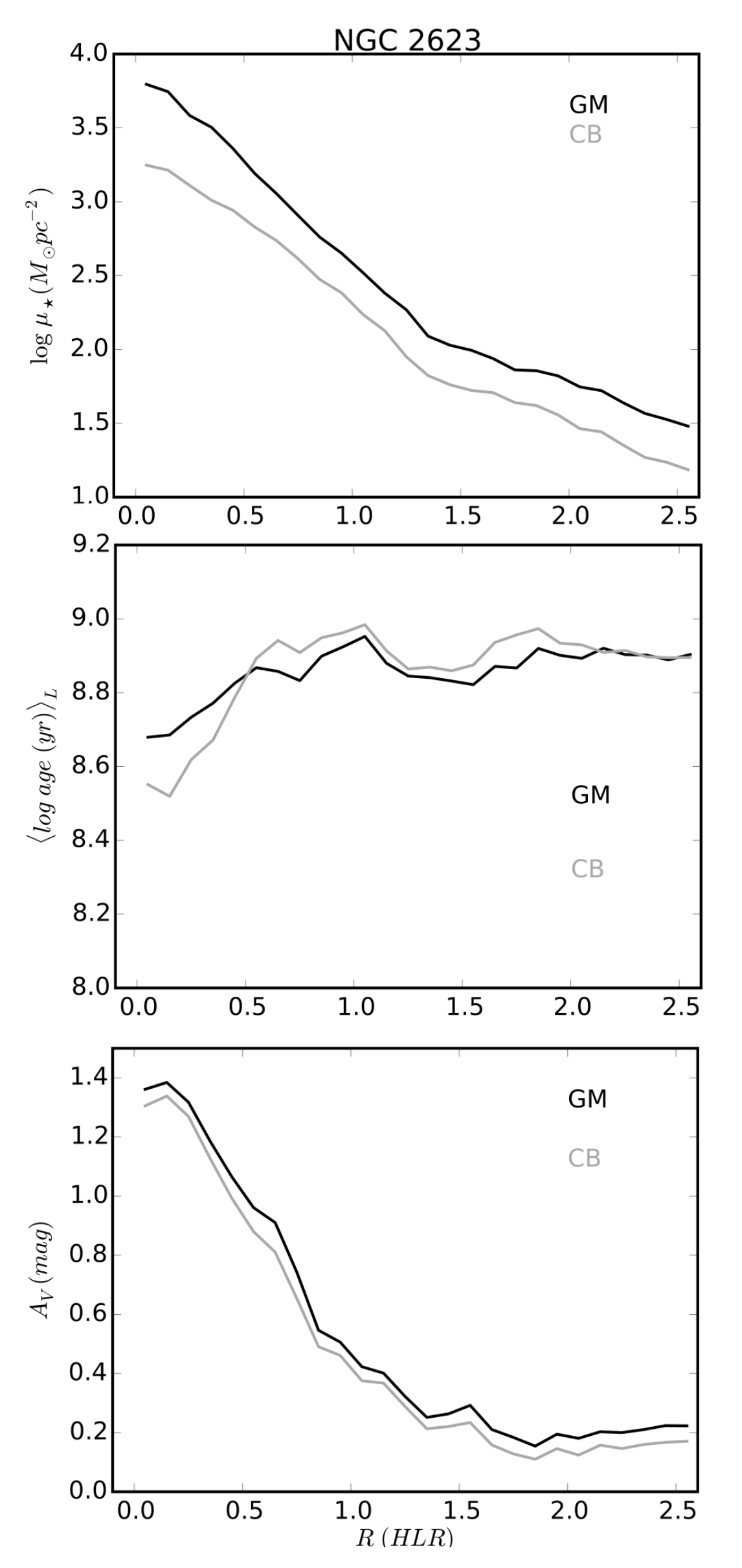}
\caption{Comparison of the radial profiles of the stellar 
population properties derived with GM base (black lines) 
and CB base (grey lines). From top to bottom: stellar mass 
surface density, mean light-weighted stellar ages, 
and stellar dust attenuation.}
\label{Fig_5p18}  
\end{center}
\end{figure} 

As mentioned in Section \ref{3.1}, we performed 
two sets of $\textsc{starlight}$ fits using the so-called 
GM and CB bases. In the paper we presented the results 
with GM base. To take into account the uncertainties 
related with the models choice, here we will compare 
the results between both bases, focusing on CALIFA data.
 
\cite{cidfernandes2014} calculate the uncertainties 
in the stellar population properties due to the choice of 
model base (GM or CB), using $\sim$100 thousand spectra from 107 galaxies 
observed by CALIFA.
They found that the uncertainties 
related to the choice of base models are larger than 
those associated with data and method, with 
one-$\sigma$ differences in $\langle log \ age\rangle_{L}$, 
A$_{V}$ of $\sim$ 0.15 and 0.08 dex, respectively.

In Figure B.1 we compare the radial profiles 
of NGC 2623 stellar population properties derived with bases
GM  (black) and CB (grey).
From the comparison of the mass surface density profiles obtained 
with the two bases (top panel), we find that GM yields higher mass 
densities by 0.32$\pm$0.08 dex, on average. This difference cannot be 
explained by the IMF change alone ($\sim$0.25 dex, \citetalias{gonzalezdelgado2015}), 
and it is also affected by the differences in the other stellar 
population properties, specially the stellar ages. 
At distances larger than 0.5 HLR, 
the average difference in mass densities, 0.28$\pm$0.03 dex, can be 
explained by the IMF change. It is in the inner $<$0.5 HLR where the 
discrepancy is significantly larger, 0.47$\pm$0.06 dex, coinciding  
with the region where the GM base starts to give older average stellar 
ages than CB, by 0.10$\pm$0.05 dex, while 
outwards of 0.5 HLR it is the opposite, 
 they are younger and more similar, -0.04$\pm$0.03 dex (middle panel). 
The extinction derived is very similar with both bases, with GM returning 
only slightly higher extinctions than CB, by 0.05$\pm$0.02 mag, across 
the whole spatial extension (bottom panel).

Thus, we conclude that when using 
different model bases (GM or CB) the  SFH derived is 
slightly different, as reflected by the change 
in stellar ages, specially in the 
inner $<$0.5 HLR. As a consequence, 
the mass surface density and mass dependent quantities, 
like the M/L ratio (see Figure C.2), are somewhat 
larger for GM than for 
CB, with the difference not being solely due to the 
IMF change from Salpeter (GM) to Chabrier (CB).

\section{Comparison between optical and 
NIR surface brightness profiles}\label{C}
\begin{figure}
\begin{center}
\includegraphics[width=0.5\textwidth]{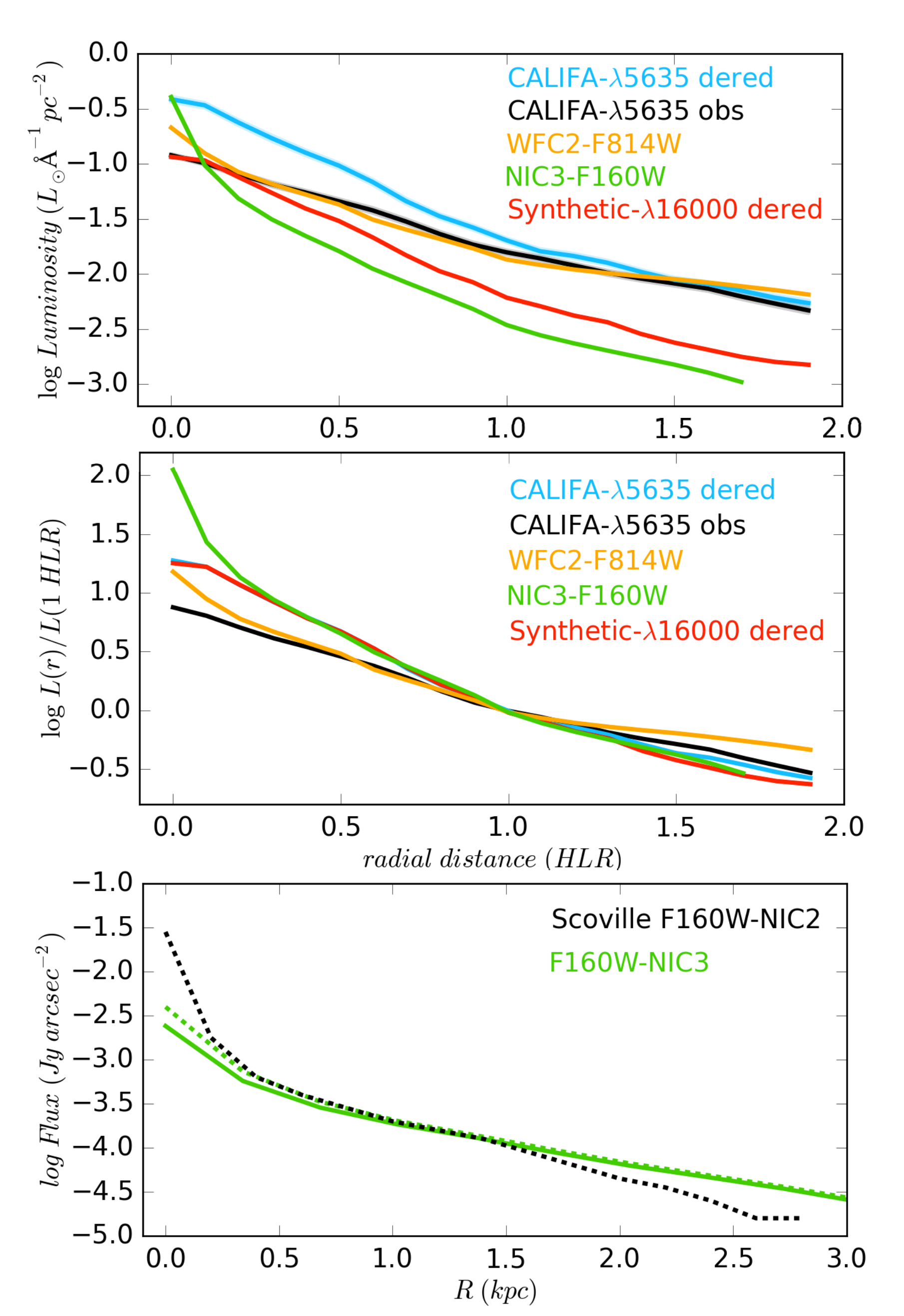}
\caption{Top panel: comparison of the different surface 
brightness profiles. Middle panel: the same profiles normalized 
at 1 HLR to compare the slopes. Bottom panel: Comparison of 
the observed (solid green line) and dereddened (dashed green 
line) NIC3 F160W brightness profile, with NIC2 F160W profile 
reported in the literature by \protect\cite{scoville2000}.}
\label{Fig_5p18}  
\end{center}
\end{figure}

In Sec. \ref{5.5} we discuss whether NGC 2623 
can evolve into an E galaxy. 
We compare the radial distribution of the 
 stellar mass surface density ($\mu_{\star}$) of NGC 2623 with 
that of non-interacting galaxies,  
previously obtained by \citetalias{gonzalezdelgado2015}. 
We find that, globally, NGC 2623 $\mu_{\star}$ is $\sim$0.6 dex less dense 
than massive E galaxies, and even $\sim$0.3 dex less dense than 
E/S0 of the same stellar mass. Also, the gradient in the central 
1 HLR is more similar to early type spirals. 
Thus, NGC 2623 is set to evolve into an Sa--Sb spiral; but we 
can not discard that it might also evolve into an E 
of $M_{\star} <$ 10$^{11}$ M$_{\odot}$. Because our estimation 
of $\mu_{\star}$ depends on the M/L ratio, and this ratio depends 
on the stellar ages and extinction, 
we need to discuss whether our estimations of these two stellar 
properties can affect our conclusion.

To check if the radial distribution of $\mu_{\star}$ 
is affected by an insufficient correction of the stellar 
extinction, we compare 
the surface luminosity profile at 1.6 $\mu$m predicted 
by our spectral synthesis to the one 
observed by NICMOS$@$HST. 
It is well known that the NIR brightness distribution 
is a good tracer of the stellar mass in galaxies, 
since it is dominated by old stars and the effect of 
extinction is lower at NIR than at optical wavelengths.

We use the HST NICMOS F160W 
image (NIC3 camera), the same that 
in the star cluster analysis in Appendix A, 
to extract the NIR surface 
brightness profile (green line in Fig. C.1) with the same geometric 
parameters as for the radial profiles of the stellar population properties  
 derived using the PPaK continuum images.
We also extract the surface brightness profile from the 
HST ACS F814W image (orange line in Fig. C.1).

From the SFH derived from the spectral synthesis, 
we have predicted the synthetic surface brightness profile 
at 1.6 $\mu$m (red line in Fig. C.1), 
using the results from the CB base\footnote{Although we 
focus on the GM base results throughout the paper, 
GM SSPs spectra only cover 
the optical range while CB SSPs spectra extent all the way 
to the NIR 1.6 $\mu$m, 
and so we use CB in this comparison.}. 
Because at each position we have the synthetic spectrum (already 
corrected by extinction), we can 
measure from it the flux at 1.6 $\mu$m and generate an 
image from which the radial profile is derived.

In the top panel of Fig. C.1 we show 
how these NIR surface 
brightness profiles compare to the observed optical surface 
brightness (black line) and to the extinction corrected optical 
surface brightness from the CALIFA datacube (blue line). 
It is clear that our "synthetic" radial profile 
at 1.6 $\mu$m is quite similar to the 
NIC3 F160W profile (a little brighter, as expected, 
since our synthetic profile is already 
corrected by extinction), 
considering the effect of the PSF in the center. 
In the middle panel 
of Fig. C.1, we show them normalized at 1 HLR in order 
to compare the slopes. We find that 
our extinction must be correct, as the slope 
of the de-reddened optical surface brightness profile 
coincides quite well with the NIC3 F160W profile, except 
in the PSF affected nuclear region.
This prevents us to trace with our data the 
inner steep rise of the de Vaucouleurs law.

We also note that the F160W brightness 
profile from NIC3 is a bit shallower than the one reported 
in the literature with NIC2 in \cite{scoville2000}. 
In the bottom panel of 
Fig. C.1 we show how they compare. The black dashed line 
is the profile by \cite{scoville2000}, the green line is 
the profile from NIC3, 
and the dashed green line is the extinction corrected profile in the 
same way as \cite{scoville2000} did in their data. We note that 
even the NIC3 profile seems to suffer from PSF smearing 
when compared to the higher resolution NIC2 profile.
Thus, we can conclude that at radial distances 
outside the nucleus our CALIFA mass/luminosity profiles are in good agreement 
with NIR estimations. 
In the nuclear regions the CALIFA data is limited by the spatial 
resolution of  PPaK  with respect to NIC2.

With respect to the stellar ages, its effect is 
already discussed in Appendix \ref{B} and shown in Fig. B.1, 
where we compare $\mu_{\star}$ 
and $\langle log \ age\rangle_{L}$ derived from GM and from CB. 
Due to the change of IMF, $\mu_{\star}$ is a 
factor 1.8 higher 
with GM than with CB. Also, due to some differences in the 
SFH derived with the two sets of 
SSPs, the stellar ages derived with GM  are 
older than those derived with CB. As a consequence, GM fits provide 
a higher M/L ratio in the central 0.5 HLR of NGC 2623, 
and a gradient $\sim$ 0.3 dex steeper. 
We note that these are intrinsic uncertainties 
associated to the analysis which we cannot break.

In anycase, we find that the M/L$_{5635}$ (GM) 
values derived by us (Fig. C.2) 
are in good agreement with the ones expected from the 
M/L$_g$ and M/L$_r$ calibrations provided by \cite{bell03} 
assuming $(g - i) = 1.1$, which is the SDSS color of NGC 2623; 
or $(g - i) = 0.86$ if the color is corrected by an average 
extinction of A$_V$ = 0.5 mag. These expressions by \cite{bell03} 
where computed assuming an exponentially declining star 
formation rate, and a Salpeter IMF. Further, our M/L$_{5635}$ results 
for NGC 2623 are also in the range of expected values for Sbc--Sc galaxies 
in the CALIFA sample (Garc\'ia-Benito et al., in preparation). 

However, the M/L at 1.6 $\mu$m, 
derived by the ratio between the radial profile 
of $\mu_\star$ (CB) and the NIC3 surface brightness profile 
at 1.6$\mu$m and between $\mu_\star$ (CB) and the 
1.6$\mu$m synthetic radial profile are a factor $\sim$4 
and $\sim$2 higher, respectively, with respect to the 
0.5 (M$_\odot$/L$_\odot$) value 
expected for a Chabrier IMF \citep{zibetti09}. The origin of this 
discrepancy is not clear for us, but its investigation 
is out of the scope of this paper.

\begin{figure}
\begin{center}
\includegraphics[width=0.5\textwidth]{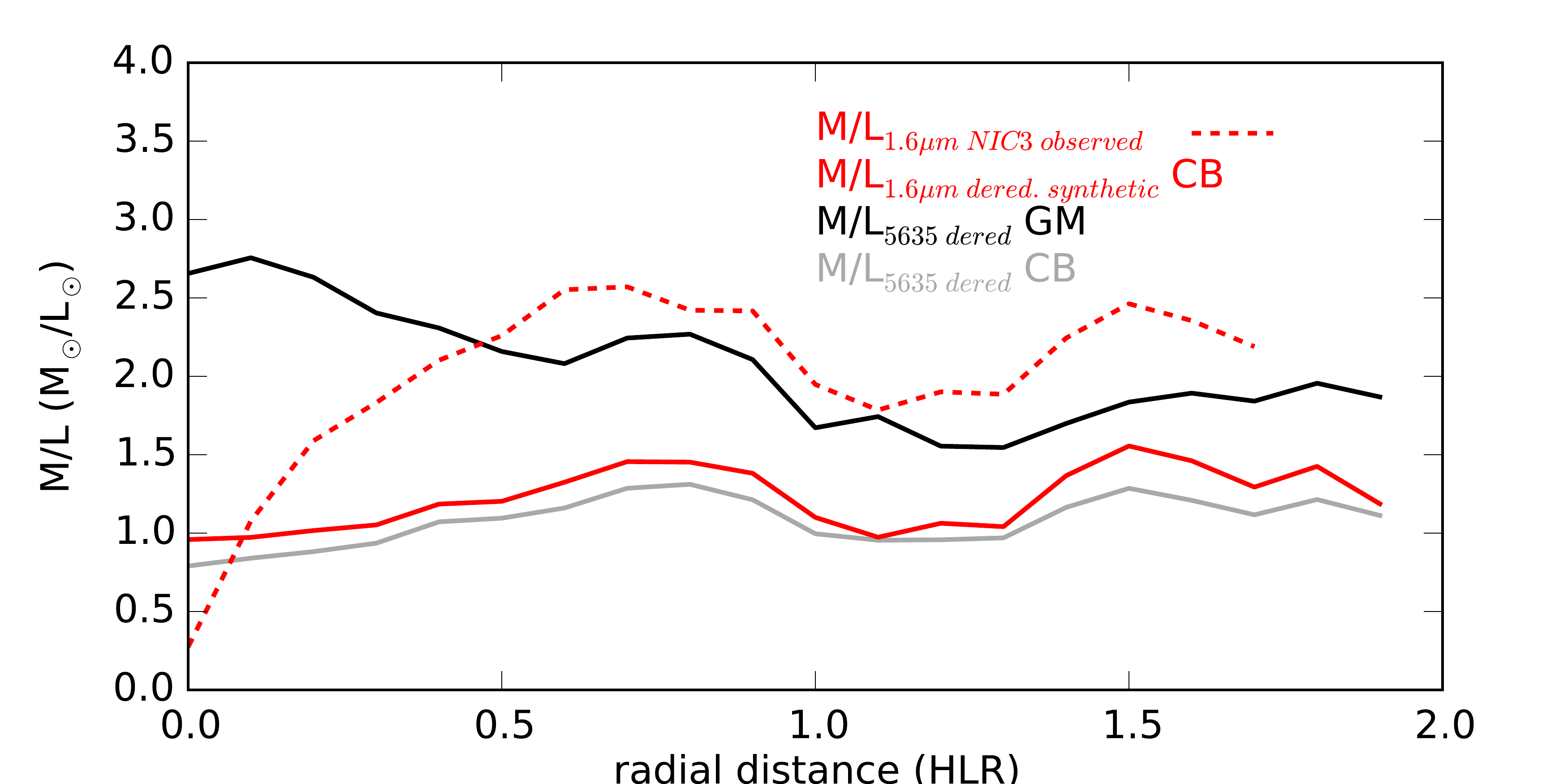}
\caption{The radial profile of M/L$_{5635}$ derived from 
the ratio between $\mu_{\star}$ and $L_{5635}$ (the luminosity 
measured at the normalized wavelength corrected by the stellar 
extinction) obtained for GM (black) and CB (grey) bases. M/L at 
the observed 1.6 $\mu$m and at the synthetic 1.6 $\mu$m are 
shown as dashed-red and solid-red lines, respectively.}
\label{Fig_c2}  
\end{center}
\end{figure} 


\end{document}